\documentclass[preprint,10pt]{elsarticle}
\usepackage{amsmath}
\usepackage{amssymb}

\usepackage{color}

\let\mpar=\marginpar
\renewcommand\marginpar[1]{\mpar{\raggedright \scriptsize #1}}

\usepackage[labelfont=bf,labelsep=period,justification=raggedright]{caption}

\makeatletter
\renewcommand{\@biblabel}[1]{\quad#1.}
\makeatother
\date{}

\usepackage{graphicx}
\usepackage{dcolumn}
\usepackage{bm}
\usepackage{hyperref}

\def\be{\begin{equation}}   \def\ee{\end{equation}}

\usepackage{xr}
\begin{document}

\begin{flushleft}
{\Large
\textbf{Membranes by the Numbers}
}
\\
\bf{Rob Phillips}$^{\ast}$
\\
 Department of Applied Physics and Division of Biology and Biological Engineering, California Institute of Technology, Pasadena, California, U.S.A
\\

$\ast$ E-mail: phillips@pboc.caltech.edu
\end{flushleft}


\section*{The Quantitative Membrane Landscape}

The pace at which  biology is advancing is staggering.
Just as there was a short 50 year gap between the invention of manned
flight by the Wright Brothers and the beginning of the space age,
in the little more than a half century since the discovery of the structure of DNA and its interpretation through the genetic code, the life sciences have entered their own age, sometimes dubbed ``the genome age''.  But there is more to living matter
than genomes.   While the genome age has unfolded, a second biological revolution has taken place more quietly.  This other success story in the emergence of modern
biology is the unprecedented and detailed microscopic view of cellular
structures that has been garnered  as a result of the emergence
of new ways to visualize cells.  Both electron
and optical microscopy have afforded an incredible
view of the  cellular interior.
In addition, the use of techniques for profiling the molecular contents
of cells has  provided a detailed, quantitative view
of the proteomes and lipidomes of both cells
and the viruses that infect them meaning that, in
broad brush strokes,  we know both what molecular components the cell is made of and
how the cellular interior looks.    A particularly fertile example
that serves as the backdrop for the present chapter is given by
our ever improving understanding of the membrane organization associated
with the organelles and plasma membranes of cells of many kinds.\cite{Buehler2016}


The goal of this chapter is to develop a feeling for membranes
in the form of biological numeracy.  That is, for the many different
ways we can think about membranes whether structurally, mechanically
or electrically, we will try to formulate those
insights in quantitative terms.
The strategy used here is to move back and forth between
a data-based presentation in which key quantitative facts
about membranes are examined,  and a rule-of-thumb and simple-estimate mentality,
in which we attempt to reason out why those numbers take
the values they do.
For those cases in which we introduce hard data, our device will
be to use the so-called BioNumbers ID (BNID).\cite{Milo2010}  Some readers will already be familiar
with the PMID (Pubmed ID) that links the vast biological literature
and databases.  Similarly, the BioNumbers database provides a curated
source of key numbers from across biology.  By simply typing the relevant
BNID into your favorite search engine, you will be directed to the
BioNumbers website where both the value of the parameter in
question will be reported as well as a detailed description from
the primary literature of how
that value was obtained.   Unfortunately, my presentation is
representative rather than encyclopedic.  There is much more
that could have (and should have) been said about the fascinating
question of membrane numeracy.  Nevertheless, the hope is that this gentle
introduction will inspire readers to undertake a more
scholarly investigation of those topics they find especially interesting,
while still providing enough quantitative insights to develop
intuition about membranes.

\begin{figure}
\centering{\includegraphics[width=4.5truein]{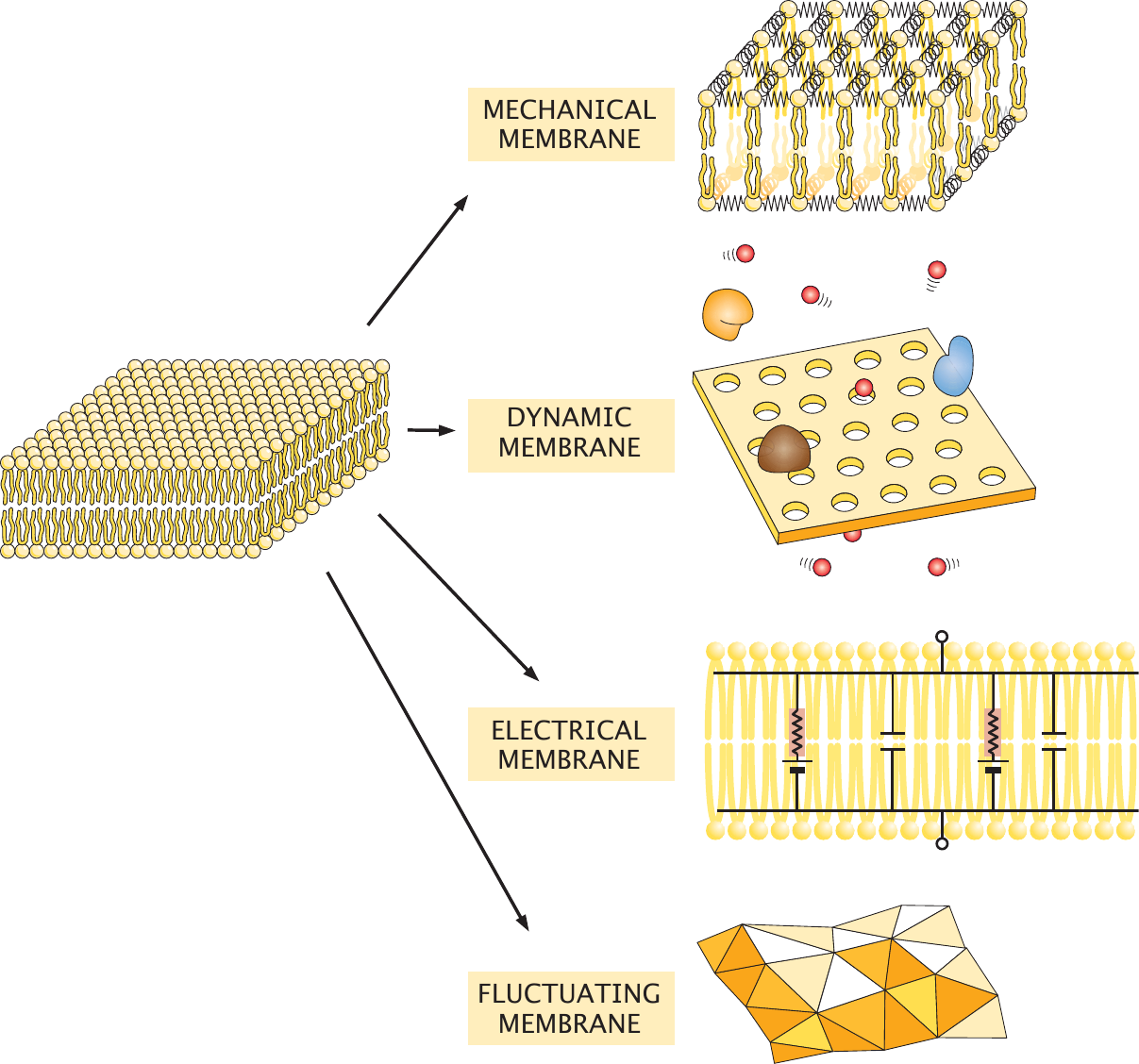}}
\caption{The many quantitative faces of a membrane.  Depending upon the
experiments being done or the questions being asked, the way we characterize
membranes is different.  When thinking about mechanical deformations of a membrane,
we will characterize it in terms of elastic constants.  Mass transport across and within membranes is described by permeability and diffusion coefficients, respectively.   When describing changes in the membrane potential, we characterize the membrane in terms of its conductivity and capacitance.   Statistical mechanics teaches us to think about
membranes from the standpoint of their fluctuations which interestingly contribute to
the membrane tension.   Each section of the chapter explores one of these
ways of characterizing membranes from the point of view of biological numeracy.
\label{fig:NumericalMembrane}}
\end{figure}

There are many conceivable organizational principles for providing biological numeracy for membranes.  The strategy to
be adopted here is to organize the numbers that
characterize membranes along several key axes,
starting with their sizes and shapes, turning then
to their chemical makeup, followed in
turn by some key themes such as the mechanics of membrane deformations, the
transport properties of various molecular species across and within membranes and the electrical properties
of membranes.  In particular, depending upon the context,
there are many different ways of thinking about membranes (see Figure~\ref{fig:NumericalMembrane}) and each of these different pictures of
a membrane has its own set of characteristic parameters.
Once these parameters are in hand,  we then attempt to make sense of all of
these numbers in a section on membrane Fermi problems with
the ambition of this section being to give an order-of-magnitude feeling
for the numbers that characterize membranes.\cite{Mahajan2014,Phillips2012} The notion of a Fermi
problem refers to the penchant of Enrico Fermi to find his way to
simple numerical estimates for complex phenomena of all kinds in short order.
The chapter closes with a look
to the future that lays out my views of some of the key challenges that
await the next generation of scientists trying to further the cause of
membrane numeracy.

\section*{The Geometrical Membrane: Size and Shape}

An inspiring  episode from the history of modern science that relates deeply
to biological numeracy  was the unfolding of our understanding of lipids and the
kinds of extended structures they make both in the laboratory and in living cells.
In his book ``Ben Franklin Stilled the Waves'', Charles Tanford gives a charming and insightful tour of this development starting
with the efforts of  Franklin who was intrigued by
the capacity of lipids when spread on water to ``still the waves''.  Indeed, this fascination led Franklin  to a famous experiment in which a spoonful of oil was seen to cover nearly half an acre of Clapham Common near London, giving a first indication of the molecular
dimensions of lipids.

Franklin's insights into the structural significance of thin films of lipids led a century later to the emergence of more formal laboratory methods for studying lipid monolayers.  In a short
1890 paper on the subject, Lord Rayleigh notes
``In view,
however, of the great interest which attaches to the determination of
molecular magnitudes, the matter seemed well worthy of investigation''.
To that end, he performed a table-top version of the Franklin experiment
concluding that for a film of olive oil he could actually compute the thickness of a monolayer, reporting a lipid length of 1.63~nm.\cite{Tanford2004}
Agnes Pockels in a letter to Lord Rayleigh published in Nature only a year later described her efforts with a trough
and force measuring balance to explore surface tension of films on water surfaces.\cite{Pockels1891}
But above all, the study of the ``determination of molecular magnitudes''
entered a new stage as a result of   a
 tour de force investigation by Irving Langmuir
 that really gave a first detailed molecular view
 of lipid molecules and the kinds of collective
 structures they can form.

\begin{figure}
\centering{\includegraphics[width=6.0truein]{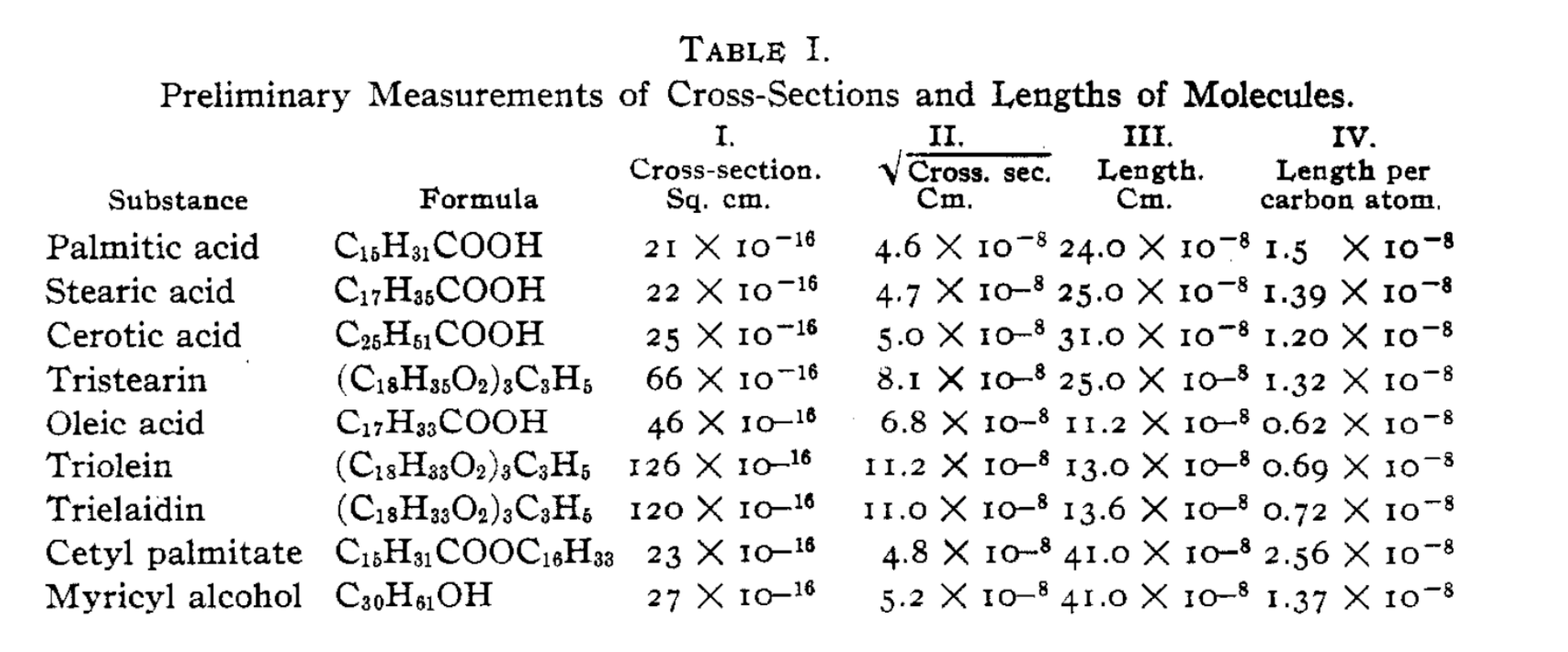}}
\caption{Lipid sizes as obtained by Irving Langmuir.\cite{Langmuir1917}  This table shows
that already a century ago, indirect methods had  yielded a
quite modern picture of lipid geometry.
\label{fig:LangmuirTable}}
\end{figure}

 Langmuir walks us through his
 experiments and deep musings about
the shape of lipids in  his paper entitled
``The Constitution and Fundamental Properties of Solids and Liquids. II. Liquids''.
Here I reproduce a lengthy but interesting series of quotes from that paper,
where Langmuir says:
 ``In order to determine the cross-sections and lengths of molecules in oil
films, experiments similar to those of Marcelin were undertaken.
The oil, or solid fat, was dissolved in freshly distilled benzene, and, by means of a calibrated dropping pipet, one or two drops of the solutions were placed upon a clean water surface in photographic tray. The maximum area covered by the film was measured.  Dividing this area by the number of molecules of oil on the surface,
the area of water covered by each molecule is readily obtained.
The results are given in the first column of Table I.''
Langmuir's Table I is
reproduced here as our own Figure~\ref{fig:LangmuirTable} and shows the impressive
outcome of his work, providing not only key numbers but also a much-needed
object lesson in the power of indirect experimental methods.
He then goes on to tell the reader how he found the lengths of these same molecules
noting,
``The volume of each molecule is found by dividing the ``molecular volume'' of the oil (M/$\rho$) by the Avogadro constant N. By dividing this volume by the cross-section of each molecule, the length of the molecule in a direction perpendicular to the surface can be obtained.''

Langmuir then goes on to say: ``It is interesting to compare these lengths with the cross-sections. As a rough approximation we may assume that the dimensions of the molecule in directions parallel to the surface can be found by taking the square root of cross-section. This is equivalent to assuming that each molecule in the surface film occupies a volume represented by a square prism with its axis vertical. The length of the square side, which we shall refer to as the average diameter, is given in the second column of Table I, while the height of the prism (or the length of the molecule) is given in the third column.'' Again,
the reader is encouraged to refer to Figure~\ref{fig:LangmuirTable} to see Langmuir's results.
He then proceeds telling us ``It is seen at once that the molecules are very much elongated. Thus
the length of the palmitic acid molecule is about 5.2 times the average diameter. The results prove that the molecules arrange themselves on the surface with their long dimension vertical as is required by the theory.''\cite{Langmuir1917}  Langmuir went much farther commenting on the significances of the
different lengths and areas emboldening him even to think about the role of unsaturated bonds in determining molecular shape.
 Indeed, one of my favorite aspects of
these experiments from Langmuir is that they led him to understand both
the number of tails and their degree of saturation truly providing a
detailed molecular picture of these molecules.
This work went even farther in the hands of Gorter and Grendel who used
similar trough experiments to hypothesize that biological membranes are lipid bilayers, a
subject we will take up again in the section on ``The Electrical Membrane'', though
I note that there are subtleties about the Gorter and Grendel approach that continue to escape me since in their analysis, they did not account in any way for the fraction of the membrane that is taken up by membrane proteins.\cite{Gorter1925}

What we see from this short historical interlude is that already at the beginning of the twentieth
century, long before tools such as x-ray diffraction and nuclear
magnetic resonance had made their way onto the scene of modern
biological science, scientists had already gleaned a detailed
view of the makeup of lipids and started to synthesize a view of how
they assemble in cell membranes.   The same story already told by experiments using Langmuir troughs has been
retold much more accurately  on the basis of x-ray and electron microscopy  experiments.\cite{TristramNagle2000,Mathai2008}
Indeed, an assessment of the current state of the art for the same
kinds of questions originally broached by Langmuir can be seen
in Table~\ref{tab:2}.

%

\begin{table*}[t]
  \centering
  \begin{tabular}{lcr}
  {\bf Lipid} & {\bf Area/lipid (nm$^2$)} & {\bf thickness (nm)} \\

DLPE  & 0.51 $\pm$ 0.005 & 2.58 \\
     DOPS  & 0.65 $\pm$ 0.005 & 3.04 \\
     DMPC & 0.61 $\pm$ 0.005 & 2.54 \\
     DLPC& 0.63 $\pm$ 0.005 & 2.09 \\
          POPC & 0.68 $\pm$  0.015 & 2.71\\
            diC22:1PC & 0.69 $\pm$  0.0005 & 3.44\\
     DOPC & 0.72 $\pm$ 0.005 & 2.68\\
  \end{tabular}
  \caption{Summary of modern version of measured lipid geometric parameters
  to be compared to those from Langmuir shown in Figure~\ref{fig:LangmuirTable}.
  All values taken from \cite{TristramNagle2000}}
  \label{tab:2}
\end{table*}

The rules of thumb that emerge
from a century of study of these molecules is that we should think of
lipid masses as being in the range of many hundreds of Daltons up to thousands of Daltons
for the largest lipids.  The lengths of these lipids vary with tail lengths of $\approx$ 2-2.5~nm.  The tail-length rule of thumb can be articulated more
precisely in terms of the number of carbons in the tail ($n_c$) as
$l_c=n_cl_{cc} $, where the length of a carbon-carbon bond is
approximately $l_{cc}\approx 0.13$~nm.\cite{Boal}   The cross sectional areas of lipids
can be captured by a rule of thumb that the area per lipid is $\approx$ 0.25 - 0.75~nm$^2$.  Note that the use of a single cross-sectional area is overly
facile because lipids can have much richer shapes than the ``square prism with its axis vertical'' described by Langmuir.  Indeed, because lipids can have shapes more like
wedges, this can lead to spontaneous curvature, a topic that we will not delve into
more deeply here, but that is critical to understanding the relation between
membrane shape and lipid geometry.
These rules of thumb are based upon a host of different measurements,
with the
thickness and area per molecule found here (BNID 101276, 104911, 105298, 105810,  105812).
We have traveled a very long way since the days of Langmuir, since
we can now order designer lipids with specific chemical properties
and even with special groups attached making these lipids fluorescently
labeled.

%



\begin{figure}
\centering{\includegraphics[width=6.0truein]{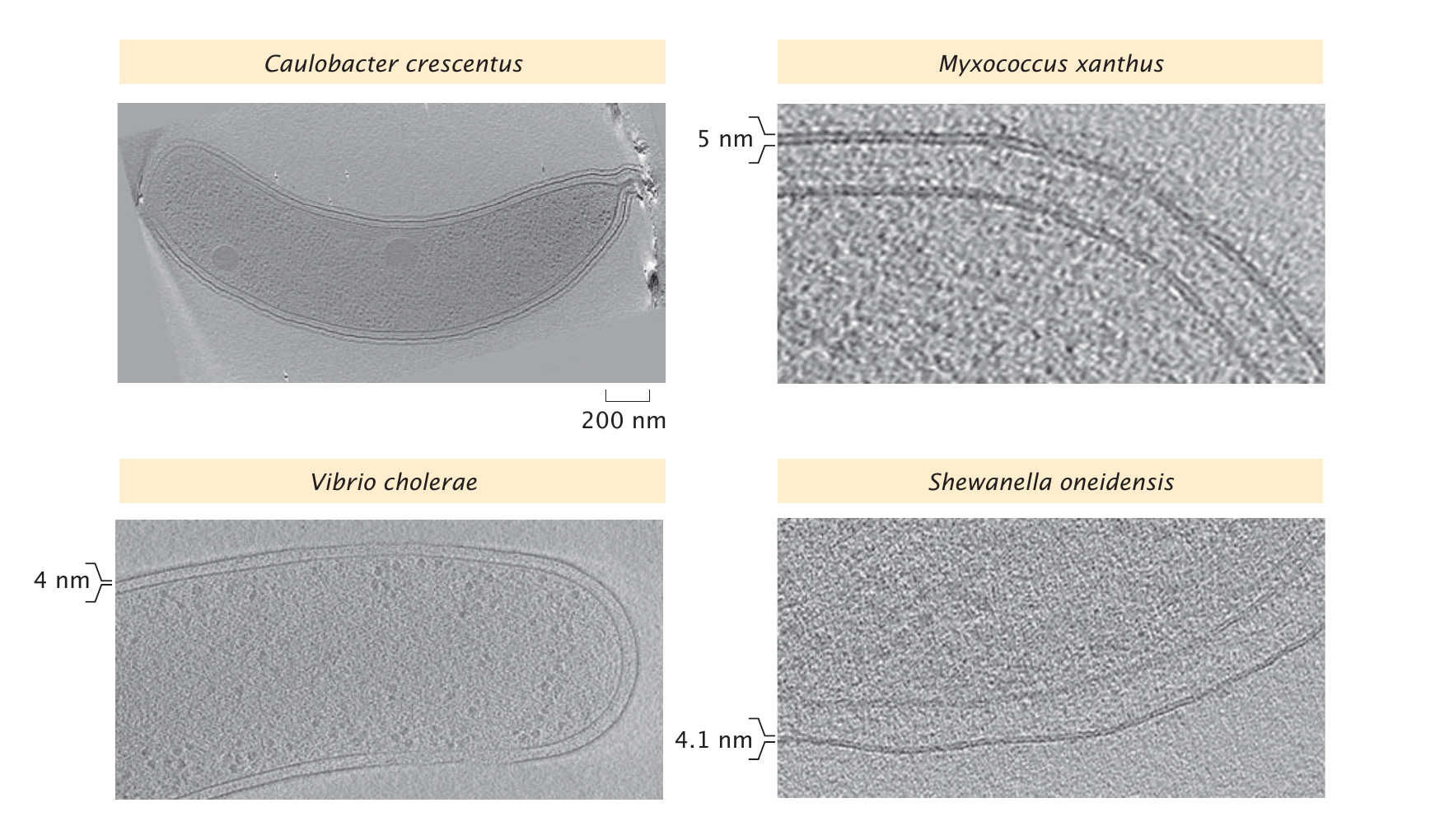}}
\caption{Electron cryo-microscopy images of bacterial cell walls.  The {\it Caulobacter
crescentus} cell gives an impression of overall cell dimensions while the higher-resolution
images of other bacteria zoom in on their membranes.  Note that these are gram-negative bacteria meaning that their external membrane architecture consists of an inner membrane, a cell wall and an outer membrane.  (Images courtesy of Grant Jensen and his laboratory members)
\label{fig:CaulobacterMembrane}}
\end{figure}

A higher-level view of the structure of cell membranes has been developing on the basis
of electron cryo-microscopy which offers an unprecedented view of the
very same structural features already explored a century ago
using the kinds of indirect methods described above.
Figure~\ref{fig:CaulobacterMembrane} provides a collage of
electron cryo-microscopy images of bacterial cell membranes.
We see that in most of these cases, the inner and outer membranes
are easily resolved and that they have a thickness of roughly 5~nm (BNID 104911).
To be more precise,
we should bear in mind that in quoting numbers such as a membrane
thickness of 5~nm, of course, we are talking about a characteristic dimension since the interaction of the lipids with the surrounding proteins can induce
thickness variations due to the effect of hydrophobic matching of the proteins
and lipids.\cite{HuangBJ1999, Nielsen2000, PhillipsUrsell2009}
Since the bacteria themselves are several microns in length
and a bit less than a micron in diameter, we can make a simple
estimate of the overall membrane area of the inner and outer
membranes by thinking of the bacterium as a spherocylinder
with a characteristic volume of $1~\mu m^3 \approx 1 fL$ and
a corresponding surface area of 5-10 $\mu m^2$.

\begin{figure}
\centering{\includegraphics[width=5.0truein]{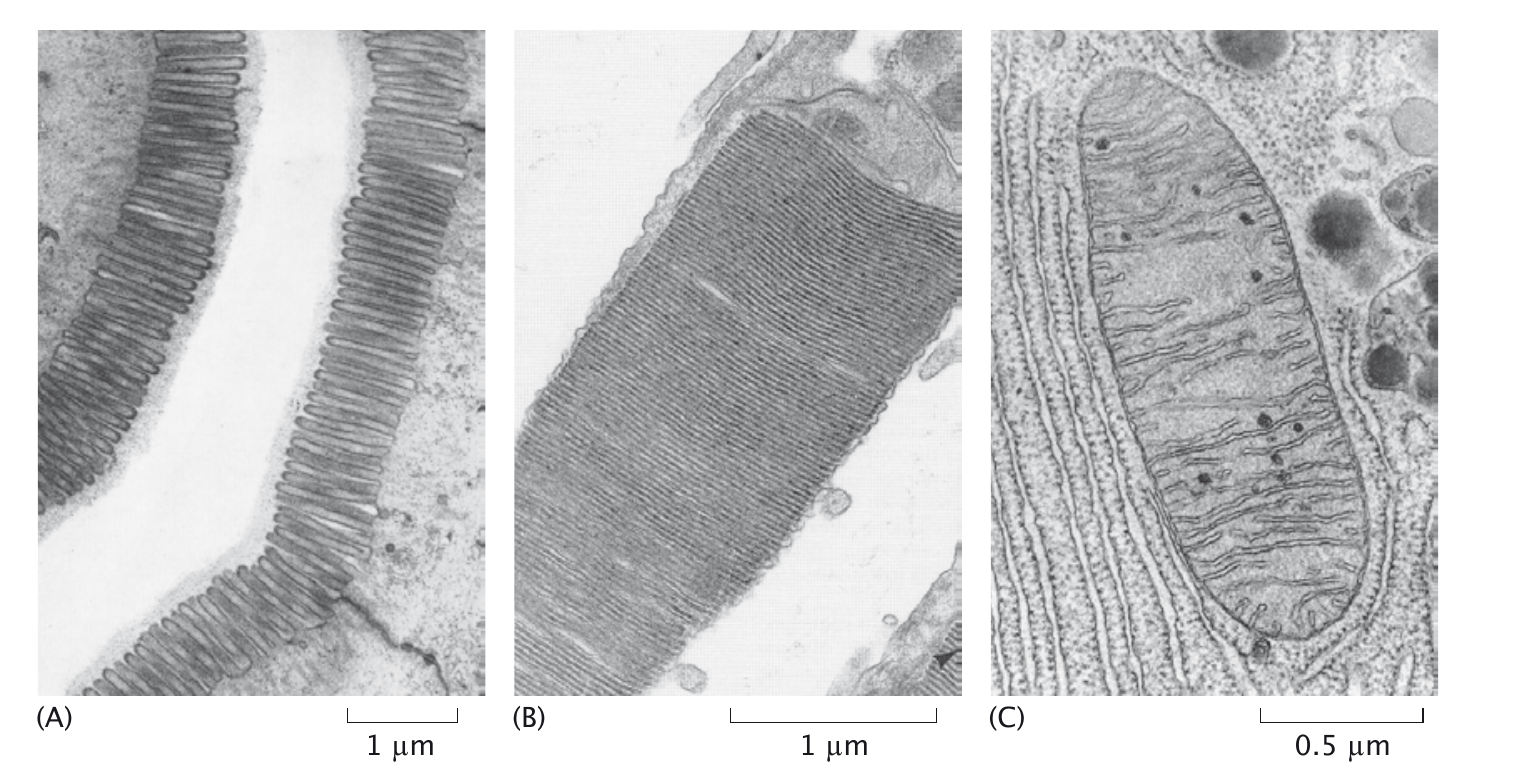}}
\caption{Eukaryotic membrane structures. (A) Apical surface of intestinal epithelial cells showing the dense membrane folds around the microvilli. The sugar chains extending outwards from the surface of the membrane can also be seen as a fuzzy layer above the microvilli. (B) Stacks of membranes packed with photoreceptors in the outer segment of a rod cell. (C) Thin section of a mitochondrion surrounded by rough endoplasmic reticulum from the pancreas of a bat. (All figures adapted from ``Physical Biology of the Cell'', Garland
Press, 2012)
\label{fig:EukaryoticOrganelles}}
\end{figure}

The membranes of eukaryotic organisms are typically more heterogeneous
and complex than those shown in Figure~\ref{fig:CaulobacterMembrane}.
Figure~\ref{fig:EukaryoticOrganelles} gives several examples coming
from electron microscopy to make that point.
First, such cells, like their prokaryotic counterparts, have an external plasma membrane
that separates them from the rest of the world.   But as seen in
Figure~\ref{fig:EukaryoticOrganelles}(A), even the cell surface can
adopt extremely complex geometries as exemplified by the microvilli.
One of my favorite examples in all of biology is shown in Figure~\ref{fig:EukaryoticOrganelles}(B) where we see the outer segment of a photoreceptor
with its dense and regular array of membrane stacks.
  However, it is perhaps the spectacular organellar
membranes (see Figure~\ref{fig:EukaryoticOrganelles}(C)) that  give a  sense of the great challenges
that remain in understanding membrane shape in cells.\cite{Neupert2012}
 Structural complexity similar to that found in
the mitochondria abounds in other organelles such as the endoplasmic
reticulum.\cite{Shibata2006, Shibata2010}

%

\begin{figure}
\centering{\includegraphics[width=5.0truein]{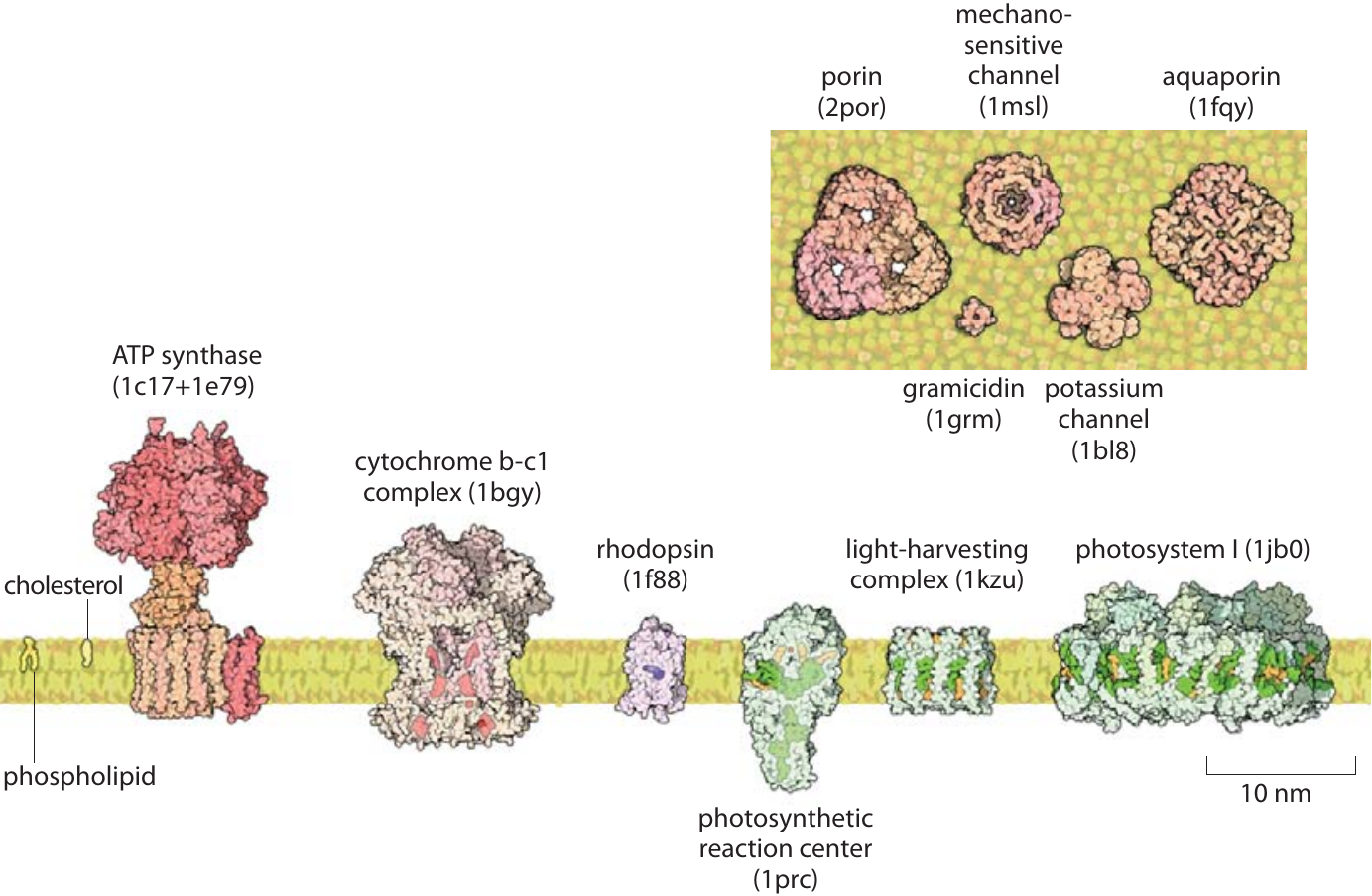}}
\caption{Sizes and shapes of membrane proteins.\cite{Milo2016}  Top and side views
of several notable membrane proteins. Note the 10~nm scale bar,
though the membrane thickness can also be used as a scale marker
as indicated in Figure~\ref{fig:CaulobacterMembrane}.
(Images courtesy of David Goodsell and adopted from ``Cell Biology by the Numbers'', Garland Press, 2015)
\label{fig:MembraneProteins}}
\end{figure}

Our brief foray into the size and shape of membranes and the molecules that make them up would of course be woefully incomplete without also
commenting briefly on the role proteins play in our modern view of biological
membrane structures.  Though early ideas about cell membranes painted a picture
of a sea of lipids dotted with membrane proteins, the modern view has turned out
to be altogether different.
``A picture is emerging in which
the membrane resembles a cobblestone pavement, with the proteins
organized in patches that are surrounded by lipidic rims, rather than
icebergs floating in a sea of lipids''.\cite{Takamori2006}
As a rule of thumb, we can think of the protein densities in bacterial membranes
as being $\sigma \approx 10^5$~proteins/$\mu$m$^2$.
 This can be used
in turn to estimate the typical center-to-center protein spacing in the cell membrane
as $d \approx \sigma^{-{1 \over 2}} \approx 3~nm$, a result that is uncomfortably
tight given that typical protein sizes are themselves $3-5$~nm as seen in
Figure~\ref{fig:MembraneProteins}.  The question of mean membrane-protein spacing
is also of great interest in the context of  organellar membranes, with a hint at what
can be expected in these cases given by a classic study
on synaptic vesicles.\cite{Takamori2006}

%
%

\section*{The Chemical Membrane}

With each passing generation, our understanding of the structures of
the cell is becoming more and more refined.  As shown in the previous section,
we have learned a huge amount about the structures of membranes
and the molecules that make them up.  But what about the specific
chemistry of these membranes?    One of the ways that
our picture of the membranes of living cells has been transformed is
through the ability to count up the molecules of different kinds,  both the lipids making up the plasma membrane and organelles and of the many
proteins that decorate these membranes.  In many ways, the development of
a census of lipid composition of membranes is an astonishing achievement
and has revealed not only that these membranes are heterogeneous, but
that the cell ``cares'' about its lipid composition.\cite{Takamori2006,Dupuy2008,VanMeer2008, Ejsing2009,Kalvodova2009,Sampaio2011,Layre2011,Carvalho2012,Klose2012, Klose2013}
Though there is still much left to be understood about precisely how cells keep track
of their membrane composition and why they ``care'', in this section of the chapter we focus on
what has been learned thus far about these chemical effects from a quantitative perspective.
For a pedagogical review, see chapter 4 of Buehler's interesting book. \cite{Buehler2016}

The same  membrane strategy used to separate the interior of cells from the extracellular
medium is also used for separating the cellular interior into a collection of membrane-bound organelles such as the nucleus, the endoplasmic reticulum, the Golgi apparatus and mitochondria. Each of these membrane systems is host to  lipids that come in different shapes, sizes and concentrations. There are  hundreds of distinct types of lipid molecules found in these membranes and, interestingly, their composition varies from one organelle to the next.  This is highly intriguing since these distinct membrane systems interact directly through intracellular trafficking by vesicles. This same heterogeneity
applies to the asymmetric plasma membrane, with different classes of lipids occupying the outer and cytosolic leaflets of the membrane (i.e. the two faces of the lipid bilayer).

Experimentally, the study of lipid diversity is a thorny problem. �Sequencing� a set of single or double bonds along a carbon backbone requires very different analytic tools than sequencing nucleotides in DNA or amino acids along proteins. Still, the �omics� revolution has hit the study of lipids too. The use of careful purification methods coupled with mass spectrometry have made inroads into the lipid composition of viral membranes, synaptic vesicles, and organellar and plasma membranes from a number of different cell types.
Indeed, the numbers in this section owe their existence in no small measure
to the maturing field of lipidomics, based in turn upon
impressive advances in mass spectrometry.  As noted above,
we remain largely in the fact-collection stage of this endeavor since
a conceptual framework that allows us to understand in
detail the whys and wherefores of lipid compositions
and how they change with growth conditions is quite immature.

\begin{figure}
\centering{\includegraphics[width=6.0truein]{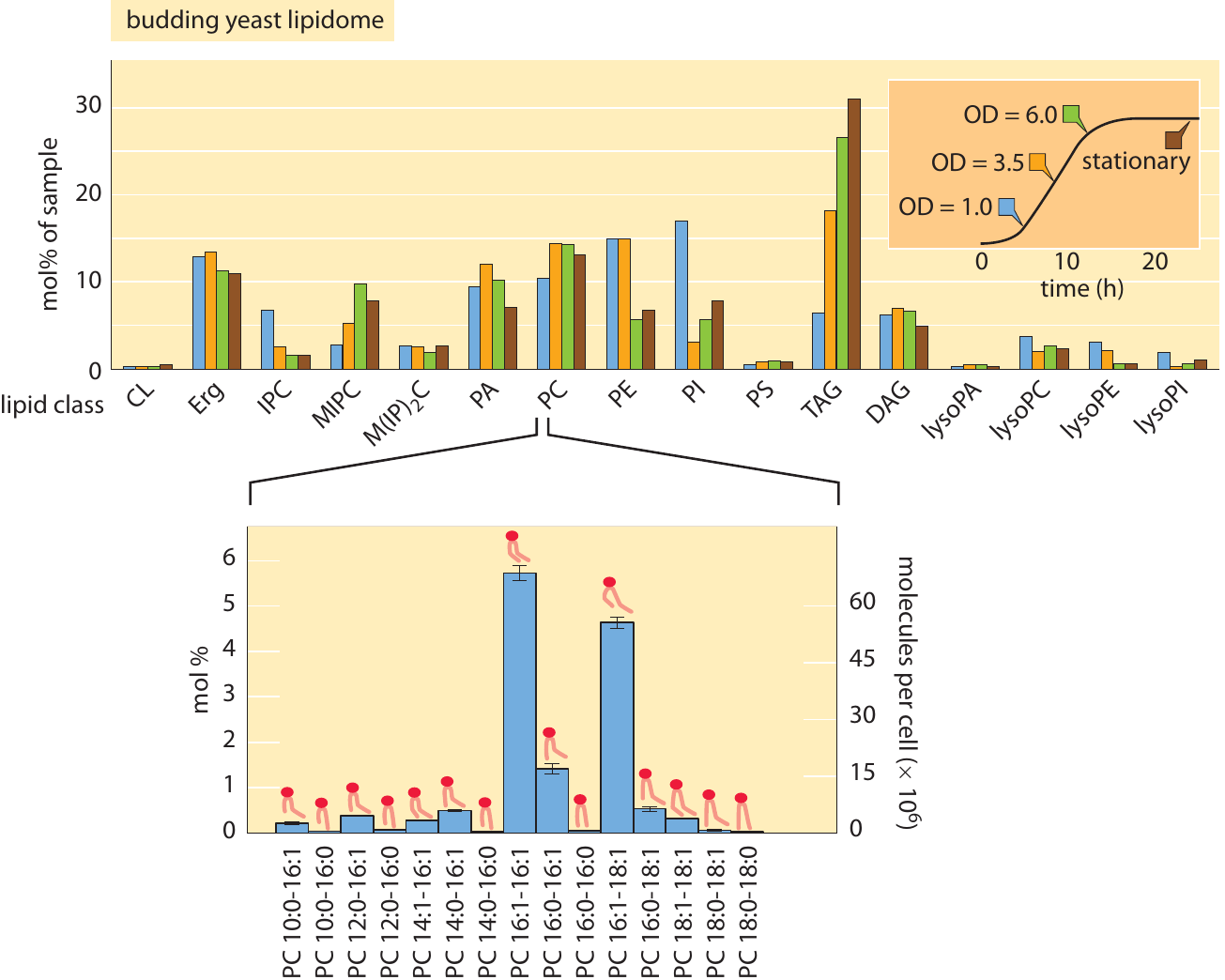}}
\caption{Lipids in yeast.  The top panel shows the relative proportions of different types
of lipid as a function of the physiological state of the cells as revealed by
the inset in the upper right.  That inset shows the result of cellular growth as measured
by spectrophotometry and leading to the optical density (OD) as a function of time. The lower panel shows the diversity of different
phospholipids.  These lipids exhibit both different tail lengths
and degree of saturation as shown by the schematics of the lipids in
the lower panel.  The abbreviations used in the
figure are:  CL: cardiolipin; Erg: Ergosterol; IPC: inositolphosphorylceramide; MIPC: mannosyl-inositol phosphorylceramide; M(IP)2C: mannosyl-di-(inositolphosphoryl) ceramide; PA: phosphatidic acid; PC: phosphatidylcholine; PE: phosphatidyl-ethanolamine; PI: phosphatidylinositol; PS: phosphatidylserine; TAG: Triacylglycerols; DAG: diacylglycerol; LPC: Lysophosphatidylcholine    Adapted from ``Cell Biology by the Numbers'', Garland Press, 2015. Data in top panel adapted from \cite{Klose2012} and data in bottom panel
adapted from \cite{Ejsing2009}.
\label{fig:YeastLipids}}
\end{figure}

Perhaps the simplest question we can pose about lipids at the outset  is how many there are in a typical cell membrane.  A naive estimate for a bacterial
cell can be obtained by noting that the area of the bacterial cell
membrane is roughly 5~$\mu$m$^2$, and recalling further that
many bacteria have both an inner and outer membrane.  To effect
the estimate, we take
\begin{equation}
N_{lipids}={4 \times \mbox{membrane area} \over \mbox{area per lipid}} \approx {20 \times 10^6 ~\mbox{nm}^2 \over 1/4 ~\mbox{nm}^2} \approx 8 \times 10^7,
\end{equation}
where the factor of 4 accounts for the fact that we have two lipid {\it bilayers}
because of the presence of both an inner and outer membrane.
This estimate is flawed, however, because we failed to account for
the fraction of the membrane area that is taken up by proteins rather
than lipids.  As was seen in the previous section on size and shape, a useful
rule of thumb is that 1/4 of the membrane area is taken up by proteins
\cite{Dupuy2008},
so our revised estimate of the number of lipids in a cell membrane would be
reduced by 25\%.
  Further, note that we used an area per lipid on the low
side and if we amended that estimate to a value of $\approx$ 0.5 nm$^2$ per lipid, this would
also bring our estimate down by a factor of two.  Literature values reported
for the bacterium {\it E. coli} claim roughly  $2 \times 10^7$ lipids per {\it E. coli} cell,
squaring embarrassingly well with our simple estimate, and leaving
us with a useful rule of thumb for the lipid density of
\begin{equation}
\sigma \approx {2 \times 10^7 ~\mbox{lipids} \over 5~\mu \mbox{m}^2 \times 4~\mbox{leaflets}} \approx 10^6 {\mbox{lipids} \over \mu \mbox{m}^2 ~\mbox{leaflet}}.
\end{equation}
\cite{Neidhardt1990}
Given our estimate of $2 \times 10^7$ lipids per bacterial cell, we can make a corresponding
estimate of the fraction of the cell's dry mass that is lipids.  As a basis for comparison, we recall
that the number of proteins per bacterial cell is $\approx 3 \times 10^6$.
\cite{Neidhardt1990,Phillips2012,Milo2016} If these proteins have an average mass of 30,000~Da, this means the total protein mass
is roughly $10^{11}$~Da or $0.15$~pg, corresponding to roughly 1/2 of the dry mass of
a bacterial cell.  For our $2 \times 10^7$ lipids, each with a mass of roughly $1000$~Da,
this means that the lipids contribute an approximate mass of $2 \times 10^{10}$~Da,
corresponding to 20\% of the protein mass, or 1/10 of the dry mass of the cell.

\begin{figure}
\centering{\includegraphics[width=4.5truein]{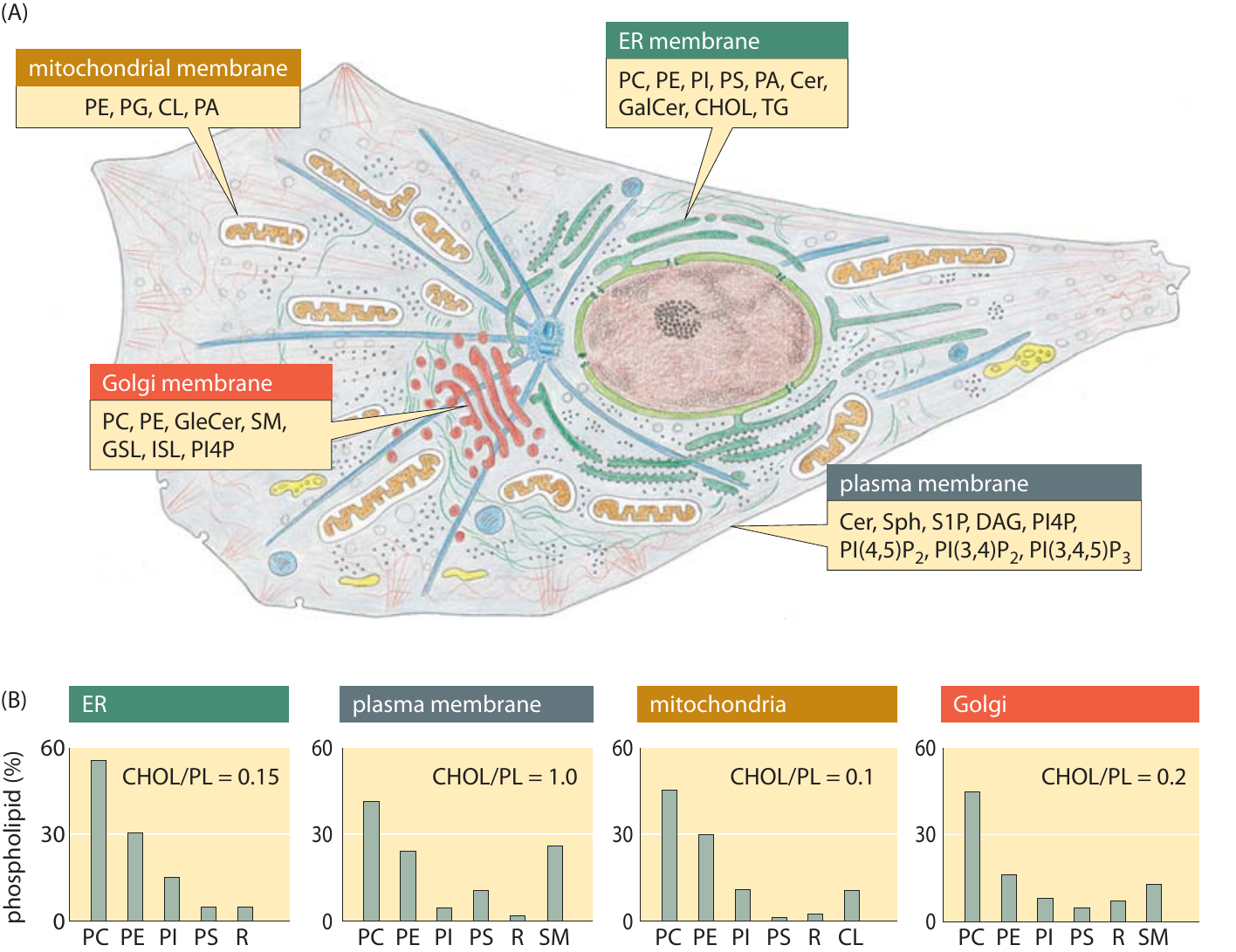}}
\caption{Organellar lipids in mammalian cells.   (A) Lipid production is spread across several organelles. The text associated with each organelle shows the site of synthesis for the major lipids. The main organelle for lipid biosynthesis is the endoplasmic reticulum (ER), which produces the bulk of the structural phospholipids and cholesterol.  (B) The lipid composition of different membranes also varies throughout the cell. The graphs show the composition out of the total phospholipid for each membrane type in a mammalian cell. As a measure of sterol content, the molar ratio of cholesterol to phospholipid is indicated. SM: sphingomyelin; R: remaining lipids. For more detailed notation see caption of Figure~\ref{fig:YeastLipids}. (Adapted from \cite{VanMeer2008})
\label{fig:OrganelleLipids}}
\end{figure}

What about the composition of membranes?  In broad brush strokes, what has been learned in lipidomic studies is that in most mammalian cells, phospholipids account for approximately 60\% of total lipids by number and sphingolipids make up another �10\%. Non-polar sterol lipids range from 0.1\% to 40\% depending on cell type and which subcellular compartment is under consideration. The primary tool for such measurements is the mass spectrometer. In the mass spectrometer each molecule is charged and then broken down, such that the masses of its components can be found and from that its overall structure reassembled. Such experiments make it possible to infer both the identities and the number of the different lipid molecules. Absolute quantification is based upon spiking the cellular sample with known amounts of different kinds of lipid standards. One difficulty following these kinds of experiments, is the challenge of finding a way to present the data such that it is actually revealing. In particular, in each class of lipids there is wide variety of tail lengths and bond saturations. Figure~\ref{fig:YeastLipids} makes this point by showing the result of a recent detailed study of the phospholipids found in budding yeast. In Figure~\ref{fig:YeastLipids}(A), we see the coarse-grained distribution of lipids over the entire class of species of lipids found while Figure~\ref{fig:YeastLipids}(B) gives a more detailed picture of the diversity even within one class of lipids.\cite{Ejsing2009}
Studies like the one presented above for yeast have also been done
in other eukaryotes as shown in Figure~\ref{fig:OrganelleLipids}.
\cite{VanMeer2008,Levental2015}
Data like this shows that the subject is even more interesting
than one might first expect because we see that lipid composition
is different for different organelles.  As noted earlier, this is especially intriguing
given the fact that these different organelles are in dynamical
contact as a result of intracellular trafficking, calling for a mechanistic and
quantitative description of how these composition heterogeneities are maintained.
All of these measurements leave us with much left to understand since as noted at the beginning of this section, the question of how cells regulate and control their lipid composition and why they care remains unanswered.

%
%
%
%
%
%
%
%

\section*{The Mechanical Membrane}

\begin{figure}
\centering{\includegraphics[width=3.5truein]{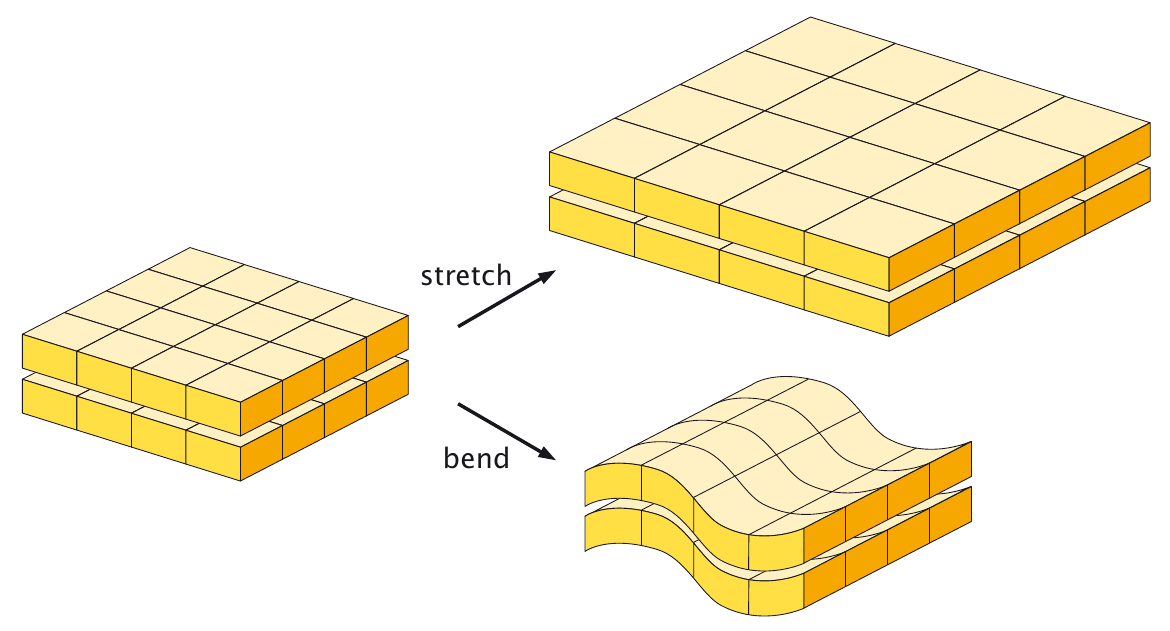}}
\caption{The mechanics of membrane deformations.  One of the deformation modes is changing the membrane area by stretching.  The second mode of membrane deformation considered here is membrane bending.
\label{fig:MembraneDeformations}}
\end{figure}

Electron microscopy images make it abundantly clear that whether
we think of the stacked membrane discs making up the outer
segment of a photoreceptor or the tortuous folds of the endoplasmic
reticulum of a pancreatic cell, biological membranes are often
severely deformed.    But as we all know from everyday experience,
changing the shape of materials usually costs energy.  As a result of membrane deformations,
energetic costs resulting from both membrane stretching and bending are incurred.   The aim of this
part of the chapter is to give a quantitative view of the energetic cost
of these deformations.\cite{EvansFaraday2013} These two different membrane
deformation mechanisms are indicated schematically in Figure~\ref{fig:MembraneDeformations}.

A natural mechanical question we might imagine starting
with is the energetic cost associated with bending the membrane.
The free energy cost to deform a tiny patch of membrane is codified in
the form of the so-called Helfrich-Canham-Evans free energy.\cite{Boal} For a tiny patch of membrane
with area  $\Delta A_{patch}$, the free energy cost to bend it is
given by
\begin{equation}
\mbox{energy to bend a membrane patch} = {\kappa_B \over 2} ({1 \over R_1}+{1 \over R_2})^2 \Delta A_{patch},
\label{eqn:Helfrich}
\end{equation}
where $\kappa_B$ is the membrane bending rigidity  and $R_1$ and $R_2$ are
the principal radii of curvature of the patch of membrane.    Note that
the  membrane bending rigidity has units of energy since the unit
of the factor in parentheses is $1/\mbox{area}$ which is cancelled by
   $\Delta A_{patch}$ which has units of area. The values of
$R_1$ and $R_2$ characterize the curvature of the surface at the point of interest.  Specifically, if we visit a particular point on the surface, we can capture the curvature by using two orthogonal circles whose radii are chosen so that those two circles most closely follow the shape of the surface at that point.   Given the free energy in eqn.~\ref{eqn:Helfrich},
we can find the total free energy of a given deformed membrane configuration by adding
up the contribution from each little patch as
\begin{equation}
E_{bend}=  {\kappa_B \over 2} \int dA({1 \over R_1(x,y)}+{1 \over R_2(x,y)})^2,
\label{eqn:Helfrich2}
\end{equation}
where now we acknowledge that the curvature (as measured by $R_1$ and $R_2$) is
potentially different at each point on the surface.
 Of course, the scale of this energy is
dictated by the bending rigidity $\kappa_B$.  Our discussion has neglected
a second topological contribution to the membrane deformation energy related
to the Gaussian curvature, though clearly such terms will be of
interest in the context of the topologically rich membrane structures found
in cell organelles.\cite{Boal}


\begin{figure}
\centering{\includegraphics[width=5.0truein]{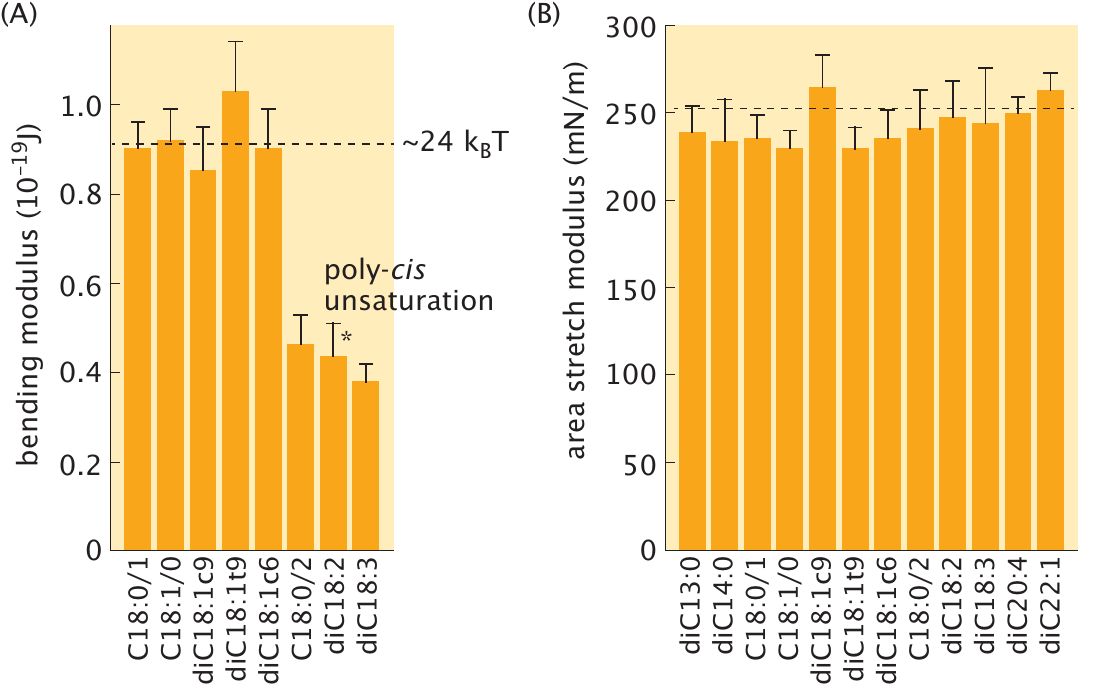}}
\caption{Elastic moduli characterizing membrane bending and stretching. (A) Values
for the membrane bending rigidity. Each value corresponds to a different lipid
with the values showing a range of tail lengths and tail saturation.  (B) Values for the area stretch modulus.
All values obtained using pipette aspiration experiments.\cite{Rawicz2000}
\label{fig:EvansModuli}}
\end{figure}

A wide range of experiments on a variety of different lipids suggest the rule of thumb that the bending modulus ($\kappa_B$)
for lipid bilayers is in the range 10-25~k$_B$T.\cite{Rawicz2000, Nagle2015}
Characteristic values of the membrane bending rigidity for
phospholipid bilayers are shown in Figure~\ref{fig:EvansModuli}(A).
We will freely
use k$_B$T for our energy units and note the conversion factors
$k_BT \approx 4.1 ~pN~nm \approx 4.1 \times 10^{-21}~J$.
The presence of sterols in lipid bilayers can increase
those numbers to $\approx 100$~k$_B$T.\cite{Song1993}
Interestingly, even measurements on biological membranes derived
from the ER and Golgi apparatus report a membrane rigidity of
$\kappa_B \approx 3 \times 10^{-19}$~J $\approx 75$~k$_B$T (BNID 110851), only a factor of three larger
than the values for phospholipid bilayers reported in Figure~\ref{fig:EvansModuli}(A).\cite{Hochmuth1996,Dai1998}

Another important question we can ask about membrane
deformations is the energy cost for
changing the area of
the membrane as seen in Figure~\ref{fig:MembraneDeformations}.
 When we stretch a membrane away from its equilibrium area, a consequence is the
 development of a tension in that membrane.
 One way to understand the magnitude
 of membrane tensions is by appealing to a so-called
 constitutive equation which loosely speaking
 relates force and membrane geometry.   In particular, the mechanics of membrane stretching is often described by
the constitutive equation
\begin{equation}
\tau=K_A {\Delta A \over A_0},
\end{equation}
where $K_A$ is the area stretch modulus  and $\Delta A$ is the area
change.  To figure out the tension, we compute the change in area, normalize
by the total unstressed area $A_0$ and then multiply by the modulus $K_A$.
In general, when we change the area of a patch of membrane by some amount $\Delta A$,
the corresponding free energy cost can be written as
\begin{equation}
\mbox{stretching energy}= {K_A \over 2}({\Delta A \over A_0})^2 A_{patch},
\end{equation}
where
 note that the units of the area stretch modulus $K_A$ are energy/area.  Several examples of the values adopted by
the area stretch modulus  are shown in Figure~\ref{fig:EvansModuli}(B),
which gives the interesting insight that for a range of tail lengths
and degrees of saturation, the area stretch modulus is nearly constant.


The actual magnitudes of the tensions in the membranes of both vesicles and cells can vary over a wide range and even the underlying mechanistic origins of these
tensions are different depending upon what regime of tension we are considering.
Interestingly, the energetics of area change is a subtle one in
the same way as the energetics of stretching a polymer like DNA is.
Specifically, let's remind ourselves of the subtleties associated
with DNA stretching as a prelude to thinking about membrane
stretching.\cite{Boal,Phillips2012} In the ``force free'' state, DNA will be
folded up and compact since such states have lower free energy
in part because the entropy of the compact conformation is higher.    To stretch DNA, the free energy cost can
be thought of as being almost
entirely entropic, meaning that with increasing stretch, there are
fewer and fewer configurations available to the DNA and hence
the entropy {\it decreases}, resulting in a net increase in free energy.
It is only when the DNA is stretched to its full contour length that we enter
a different regime that actually involves molecular bond stretching.
Because the mechanisms in these regimes are different, it should
not surprise us that they are actually characterized by different mechanical
stiffnesses.  Similar intuition emerges for the membrane case.

By analogy with polymer stretching, we can think of the energetic cost associated with
membrane deformations in much the same way.  That is,
for a floppy (low tension) membrane, stretching the membrane
has an associated free energy cost that results from
``pulling out the wrinkles'', and is effectively entropic.\cite{Boal}  At higher
tensions, the actual bond stretching effect intervenes.  Though
very few systematic insights have been obtained for thinking
about the membranes within cells, a series of rigorous, systematic
studies in lipid bilayers have set the standard in the field.\cite{Rawicz2000}
At even higher tensions, lipid bilayer membranes will actually
rupture with the rupture tensions occurring between 5~mN/m and 10~mN/m
depending upon the type of lipids in question.\cite{Olbrich2000}

Though there are fewer systematic measurements for
cellular membranes, some clever experiments have shed light
on this topic as well.   The tension measured in ER membrane networks has a value of $1.3 \times 10^{-2}$~mN/m
while that measured in the Golgi membrane is given by $0.5 \times 10^{-2}$~mN/m.\cite{Upadhyaya2004}
These numbers are quite small as can be seen by comparing them
to the membrane rupture tension which is a thousand times
larger with a range of $5-10$~mN/m as noted
above.\cite{Olbrich2000}
   Note also that the subject
of membrane tension is a tricky one in the cellular setting because
measured tensions have many contributions including from the underlying
cytoskeleton and the battery of molecular motors associated with it.\cite{Peukes2014}
There is  an excellent review
featuring both a clear discussion of the different methods as well as the range of
measured tensions.\cite{Sens2015}   Table 1 of that review includes an exhaustive listing
of measured membrane tensions as well as the caveats associated with each such
measurement.

\section*{The Dynamic Membrane}

Perhaps the defining feature of biological membranes is that they
serve as barriers between some compartment of interest (the cytoplasm,
the Golgi apparatus, the nucleus, the endoplasmic reticulum, etc) and
the rest of the world.   The very word "barrier" points toward underlying molecular rules that determine the rate at which molecules cross through or move within membranes, and thereby regulate how a cell distinguishes itself from the environment.
 In this section, we begin by
exploring the permeability of biological membranes to various molecular species.
After that, we then turn to the diffusive properties of molecules within the membrane.

One of the key ways we characterize membrane permeability is to ask
the question of how many molecules cross a given area of membrane
each second, a quantity defined as the flux, $j$.  In particular if
we have a difference in concentration  of some species across the membrane given
by $\Delta c$, then in the simplest model the flux is given by
\begin{equation}
j=-p \Delta c,
\label{eqn:Permeability}
\end{equation}
where the parameter $p$ is the permeability of interest here.
Note that a more rigorous treatment of the flux invokes the chemical
potential difference across the membrane, though for our purposes
this simple linearization suffices.\cite{kedem:1958,Manning1968}
The units of the permeability can be deduced by noting first that
the units of $j$ are
\begin{equation}
\mbox{units of} ~j = {\mbox{number of molecules} \over L^2 T}.
\end{equation}
Here we adopt the standard strategy when examining units
of physical quantities of using the symbol $L$ to signify units of
length and $T$ to signify the units of time.\cite{Robinett2015} Given
these conventions, the units of concentration are
\begin{equation}
\mbox{units of}~c = {\mbox{number of molecules} \over L^3}.
\end{equation}
The requirement that the units on the two sides of the equation balance implies
that the units of the permeability itself are
\begin{equation}
\mbox{units of} ~p = {{\mbox{number of molecules} \over L^2 T}
 \over {\mbox{number of molecules} \over L^3}} = {L \over T}.
 \end{equation}
 In the remainder of the paper, we will report units of permeability
 in nm/s, though often one finds values reported in cm/s as well.

\begin{figure}
\centering{\includegraphics[width=5.0truein]{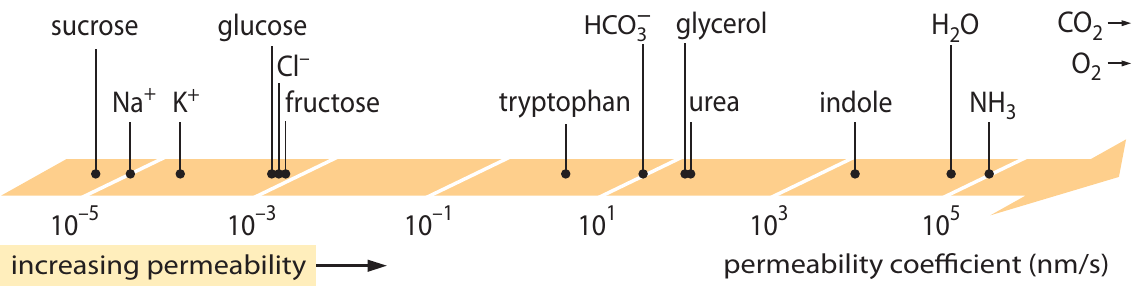}}
\caption{Range of membrane permeabilities.   Permeability coefficients
for a number of different lipid species showing the huge dynamic
range in permeability.
\label{fig:Permeability}}
\end{figure}

The first and probably most important thing we should say about
the numerical values adopted by membrane permeability is that there is no such thing as {\it the}
membrane permeability.  That is, the rate at which molecules
pass across membranes is an extremely sensitive function of which molecules
we are discussing as well as the type of molecules making up
the membrane itself.\cite{Finkelstein1976,Olbrich2000,Mathai2008}
Figure~\ref{fig:Permeability} makes this point clear by reporting the range of values
for permeability for a number of different molecular species revealing a more than
10-order-of-magnitude range of permeabilities, with the membrane
being effectively impermeable to ions such as Na$^+$ and
K$^+$, while for water molecules, the permeability is ten orders of magnitude
larger.  Though this doesn't rival the 30 order of magnitude range that is found
for electrical conductivities of different materials, these numbers still imply
a huge difference in the transport properties of different molecules across membranes.

\begin{figure}
\centering{\includegraphics[width=2.0truein]{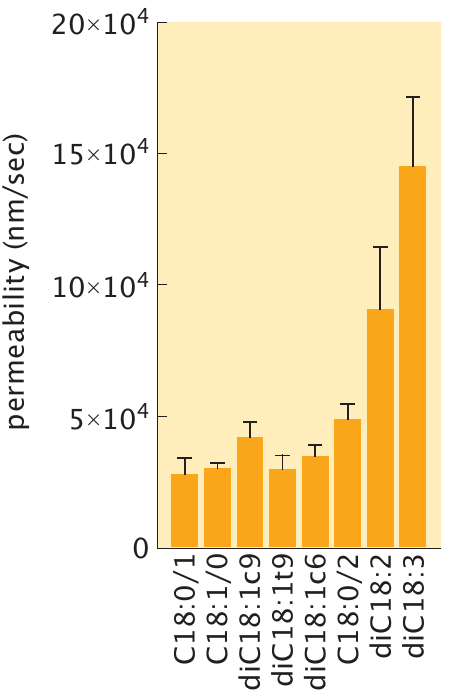}}
\caption{Range of membrane permeabilities for water. Measurements
made at 21 $^{\circ}$ C using the micropipette aspiration technique
in conjunction with video microscopy to monitor vesicle size.\cite{Olbrich2000}
\label{fig:PermeabilityMeasure}}
\end{figure}

How are such permeabilities measured?  One approach to measuring these membrane permeabilities is the use of radioactive tracer molecules.  By setting up a membrane separating two aqueous regions with different compositions, one can measure the accumulation of the tracer in one region as a result of flux from the other region over time.\cite{Finkelstein1976}  A second important set of measurements for water permeability
were performed using giant unilamellar vesicles using the so-called micropipette
aspiration method where an osmotic pressure is applied across the membrane
and the resulting flux of water across the membrane is measured.  Here the idea is that a pipette with a characteristic diameter of
several microns is used to grab onto a vesicle with a diameter of roughly
10 $\mu$m or larger.  By applying a suction pressure, the tension of the
vesicle can be monitored.  Further, by using video microscopy, the volume of the vesicle can be carefully monitored, giving a sense of the rate at which the vesicle is inflated
as a result of mass transport of water across the membrane.  The results of such measurements
for a set of different lipid types are shown in Figure~\ref{fig:PermeabilityMeasure},
with values entirely consistent with those shown schematically
in Figure~\ref{fig:Permeability}.

The classic work of Hodgkin and Huxley offered many important insights.
To my mind, one of the most interesting arguments that they made
is a testament to the role of clear theoretical (and quantitative)
thinking in biology.  In particular, they argued that the membrane permeability
to ions such as Na$^+$ and
K$^+$ must  change transiently and substantially to permit
key ions across the otherwise impermeable membrane (see
Figure~\ref{fig:Permeability} to get a sense of the extremely low permeability of
charged ions).  Specifically, they introduced a highly
nonlinear permeability response that suggested that there must
be molecules in the membrane of the cell that could selectively
change the permeability in response to changes in driving forces such
as the membrane potential, effectively hypothesizing the existence of
ion channels before they were known.

We now know that biological membranes
are littered with batteries of different channels and pumps whose job
it is to transiently alter the permeability of the membrane or to actively transport
molecular species across it.
These membrane proteins are responsible for many physiologically important functions including
the transport of ions and  sugars such as glucose  and lactose
that are critical to the cellular economy.   Ions typically pass across ion channels  at rates between $10^7$ and $10^9$ ions per second, though of course
this rate depends upon the concentration difference across the membrane itself
(BNID 103163,103164).  Glucose transporters have a much lower characteristic
rate of several hundred sugars per second (BNID 102931, 103160) while bacterial lactose transporters
have a characteristic rate of 20-50 sugars per second (BNID 103159).
Though here we report on the rates associated with several well-known membrane
proteins, more generally,
  the rates at which the various membrane proteins that are responsible
for transport operate are not that well known, with a dearth of modern data
 spanning the range of different membrane transporters (BNID 103160).\cite{Stein1990}


A second kind of  membrane
dynamics different from
the transport {\it across} the membrane
described above is diffusion of molecules laterally within the membrane.   As already noted throughout the chapter,
the membrane is a highly heterogeneous composite of lipids and
proteins and when thinking about the diffusive dynamics within
the membrane, we need to do so on a molecule-by-molecule basis.
Since we are thinking about membranes, the first class of molecules
we might be interested in characterizing are the lipids themselves.\cite{Schlessinger1977,Alecio1982,Gaede2003, Doeven2005,Nenninger2014}
For example, in eukaryotic cell membranes, by using the clever method of fluorescence-recovery-after-photobleaching (FRAP), a lipid diffusion constant of $0.9$ $\mu$m$^2$/s
was measured.\cite{Schlessinger1977}  This diffusion
constant is roughly ten-fold lower than the values that would be found in a model
lipid bilayer membrane.\cite{Fahey1977} More recent measurements
confirm these classic numbers (see Figure~4 of ref.~\cite{Nenninger2014}, for
example).

\begin{figure}
\centering{\includegraphics[width=5.5truein]{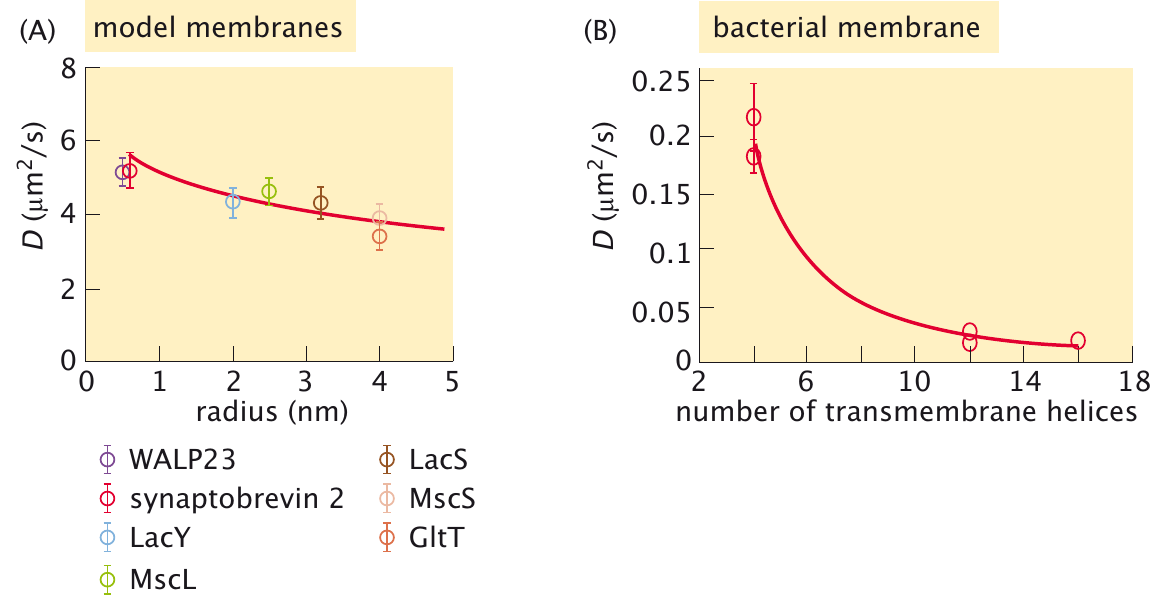}}
\caption{Range of diffusion coefficients.   (A) Diffusion coefficients for different membrane proteins measured using fluorescence correlation spectroscopy  in giant unilamellar vesicles showing dependence on
protein size.  The red line is a fit using the Saffman-Delbr{\"u}ck model which characterizes membrane diffusion as a function of the size of the diffusing molecule.\cite{Saffman1975,
Gambin2006}  (B) Diffusion coefficients for different membrane proteins
measured using fluorescence
recovery after photobleaching (FRAP) in the {\it E. coli} cell membrane. The red line is
an empirical fit as a function of the number of transmembrane helices in the protein. The names refer to particular membrane proteins used in the experiments.
(A) adapted from \cite{Ramadurai2009} and (B) adapted from \cite{Kumar2010}.
\label{fig:DiffusionMeasure}}
\end{figure}

It is of great interest to characterize the in-plane diffusion not only
of the lipids themselves, but also of the proteins that populate
those membranes.  Figure~\ref{fig:DiffusionMeasure} gives examples of
membrane diffusion constants for several different membrane proteins.
Further, we need to acknowledge the large differences in lateral diffusion
coefficients between model membranes such as are found in
giant unilamellar vesicles where the  values of diffusion coefficients for
membrane proteins are 1-10~$\mu m^2/s$ \cite{Ramadurai2009} and those
in native membranes where membrane proteins are characterized
by diffusion coefficients that are several of orders of magnitude lower
with values of 0.01-0.1~$\mu m^2/s$.\cite{Kumar2010,Chow2012,Mika2014, Nenninger2014}
However, these measurements are more nuanced than first meets
the eye and the results for several membrane proteins have been
shown to depend upon the time scales over which the diffusion
is characterized.\cite{Chow2012}  In particular, using the FCS method which
probes diffusion on short length and  time scales, both the TAR receptor and TetA (a
tetracycline antiporter) were found to have diffusion constants of
4.2~$\mu$m$^2$/s and 9.1~$\mu$m$^2$/s, respectively, to be contrasted
with the values of 0.017~$\mu$m$^2$/s and 0.086~$\mu$m$^2$/s, respectively found
when using the FRAP measurement.
Indeed, as we will note in the final section of the chapter, the question
of how best to move from biological numeracy in model membranes to biological membranes
with their full complexity  is
one of the key challenges of the coming years of membrane research.


%

%

\section*{The Electrical Membrane}

A membrane has many different properties as shown in Figure~\ref{fig:NumericalMembrane}.
So far, our picture of membranes has focused on their mechanical
and transport properties.  However, our discussion of action potentials
and the pathbreaking work of Hodgkin and Huxley already hinted
at the view that membranes can also be thought of as circuit elements.
Specifically, part of this chapter's very business is to illustrate
some of the different {\it abstract} ways of describing membranes
and what {\it effective} parameters to attribute to them.  We now jettison
the view of a membrane as a mechanical object, instead
focusing on it as a collection of resistors and capacitors
as shown in one of the panels of Figure~\ref{fig:NumericalMembrane}.


The picture already developed under the heading of the ``Electrical Membrane''
in Figure~\ref{fig:NumericalMembrane} tells us that in the presence of an electric
potential, a lipid bilayer behaves as an array of resistors and capacitors
in parallel.
One way to measure the electrical conductance across a membrane patch is to
form
a lipid bilayer membrane across a hole separating
two solutions.  Then, different voltages are applied across the membrane and
the current-voltage characteristics are measured, with the membrane
conductance then determined by using the slope of these current-voltage
curves. In our discussion of the electrical membrane we characterize electrical properties on a per unit area of membrane basis.
For the conductance, a series of measurements like those described above for a number of different charged species result
in a range of values for the bare membrane conductance of roughly
1-5 nS/cm$^2$.\cite{Hanai1965,Vorobyov2014}  To get a sense of
how small the membrane conductance is, note that if we consider a characteristic
conductance of 1~nS for an ion channel such as the mechanosensitive channels
found in bacteria \cite{Haswell2011}, if we normalize by the area this means that
the channel conductance is more than ten orders of magnitude larger than
that of the membrane itself.

But membranes have more electrical properties than their
conductance alone.\cite{Phillips2012}
Capacitance is a measure of the ability of a circuit element to store
charge.
A local disruption of charge
neutrality is permitted near surfaces. In particular, in this
setting, the capacitance is defined as  the ratio of the excess
charge on either side of the membrane and the membrane potential, $C
= q/V_{mem}$. The capacitance of a patch of the cell membrane can be
approximated by thinking of it as a parallel plate capacitor. The
charge on the capacitor plates is ${\pm}\sigma A_{patch}$, where
$\sigma$ is the excess charge per unit area of membrane, and
$A_{patch}$ is the area. The electric field inside a parallel plate
capacitor is uniform and equal to $\sigma/(\varepsilon_0 D)$, where
$D$ is the dielectric constant of the material between the plates.
Therefore the potential drop across the
membrane is $V_{mem}=\sigma d/\varepsilon_0 D$, where $d$ is the
thickness of the membrane, or the distance between the plates of the
parallel plate capacitor. Dividing the charge by the membrane voltage
leads to the formula, $C=\varepsilon_0 D A_{patch}/d$, for the
capacitance of a patch of membrane. Since the cell membrane has a
thickness of $d\approx 5$\,nm and a dielectric constant $D_{mem}=2$,
its capacitance is predicted to be $C_{area}=C/A_{patch}\approx
0.4\,\mu$F/cm$^2$. The typical measured value for the capacitance
per unit area in cell membranes  is $C_{area}=1\,\mu$F/cm$^2$.\cite{Fricke1925,Curtis1938,Almers1978}

We have already discussed the century long quest to understand the size of
lipid molecules and the membranes they make up.  We learned that one branch of these investigations
passed through the enormously impressive work of Pockels, Rayleigh and
Langmuir.  Amazingly, a completely independent line of enquiry in the hands
of Fricke related
to the electrical properties of membranes led to nearly the same
result.\cite{Fricke1925}
Using these ideas, we can recast the measured value of the membrane
capacitance as a result for the membrane thickness as
\begin{equation}
d={\epsilon \over (C/A)}= {2 \epsilon_0 \over (C/A)}\approx {2 \times 8.8 \times 10^{-12} {F \over m} \over 0.4 \times 10^{-2} {F \over m^2}} \approx 4 ~nm,
\end{equation}
a beautiful result astonishingly close to the value obtained using
the equation of state of monolayers by Pockels, Rayleigh and Langmuir.
Note that to obtain this result, we rewrote the conventional membrane capacitance of
$0.4~\mu F/cm^2$ in the more appropriate SI units as $0.4 \times 10^{-2}~F/m^2$.
Further, whereas Frick used a relative dielectric constant of 3, the estimate
used here is based upon the value of 2.
In light of the measurement of the membrane capacitance, scientists such as
Fricke realized that this would provide yet another sanity
check on the membrane thickness.\cite{Fricke1925}
In this era where many  scientists seem almost to have scorn
for the idea of figuring things out without seeing them directly,
the determination of the thickness of lipid bilayers long before the advent of
direct techniques such as electron microscopy should give readers
pause before casually dismissing results that come from
indirect measurements.



\section*{The Fermi Membrane: Thinking up  membranes}

So far, this chapter has been an ode to biological numeracy
in the context of membranes, showing us the many different
ways in which we can quantitatively describe our hard-earned
understanding of these fascinating structures.
These numbers are summarized in Table~\ref{tab:1}.
But in the abstract, such numbers are often boring and sometimes
useless, or worse yet, misleading.  To my mind,  numbers
 that characterize the world around us
are  only really interesting when put in the context
of some argument or reflection.
 For example, we know that
if we drop an object near the surface of the Earth, in the first second,
it will fall roughly 5 meters.  So what?  In the powerful hands of Newton,
 this innocuous number became part of his inference of
the law of universal gravitation.   There is a direct intellectual line
from a knowledge of the radius of the Earth and the distance to the moon
to Newton's estimate leading him to further trust the idea
that the force of gravity falls off as the square of the distance.  In that case,
he realized that the distance to the moon is roughly 60 times larger
than the radius of the Earth, meaning that the acceleration of the moon as it ``falls'' towards the center of the Earth should be
$(60)^2$ = 3600 times smaller  than that associated with that apocryphal apple falling from Newton's tree.  To finish off his estimate, he asked the question of how
far the moon falls compared to how far the apple falls when watched for the
same time and found them to ``answer pretty nearly'', with the moon falling
roughly 1/3600 as much in one second as the 5~m a falling body at the surface
of the Earth falls in that same time interval.   But what does this have
to do with our quantitative musings on membranes?  To my mind, it illustrates
how powerful simple numerical arguments can sometimes be to help us
see whether our way of thinking is consonant with the known facts about
a system.

\begin{table*}[t]
  \centering
  \begin{tabular}{lcr}
  \medskip
  {\bf membrane parameter} & {\bf range of parameter values} & {\bf BNID} \\

    lipid length  & $\approx$ 2.5 - 3.5 ~nm & see Table 1 \\
     lipid area  & $\approx$ 1/4 - 3/4~nm$^2$ & see Table 1 \\
     number of lipids per cell (bacterium) & $\approx 2 \times 10^7$ & 100071\\
     bending rigidity & $10-25$~k$_B$T & 105297\\
     area stretch modulus & 200-250 mN/m (or $\approx$ 50~k$_B$T/nm$^2$) & 	112590,
112659 \\
     membrane tension &  $10^{-4}-1~k_BT/nm^2$ &110849,112509, 112519 \\
     rupture tension & 1-2 ~$k_BT/nm^2$ & 112489, 110911 \\
       membrane permeability (water) & 10-50~$\mu$m/s & 112488 \\
       membrane capacitance & $\approx 1 \mu F/cm^2$ &	110759,	109244, 110802 \\
       membrane resistance & $ 0.1- 1.5 \times 10^9$ $\Omega$ cm$^2$  & 110802 \\
       membrane potential & 100~mV & 109775, 107759\\
       diffusion constant (lipid) & $\approx$ 1~$\mu$m$^2$/s & 112471,
112472 \\
         diffusion constant (membrane protein) & $\approx$ 0.02 - 0.2~$\mu$m$^2$/s  & 107986 \\
  \end{tabular}
  \caption{Membranes by the numbers. A summary of the key numbers about
  membranes discussed throughout the chapter/ for easy reference.  Numbers reported are ``typical'' values and should be used as a rule of thumb.  For a more detailed
  description of parameter values, the reader should use the Bionumbers database
  through the relevant BNID.  Also see Box 1 of \cite{PhillipsUrsell2009}.}
  \label{tab:1}
\end{table*}

Inspired by the long tradition of simple estimates when faced with
numerical magnitudes to describe the world around us, we now examine
the ways in which the numbers presented throughout the chapter
can help us to better understand membranes and the biological
processes that take place at them.  Indeed, we are inspired by the
notion of the so-called Fermi problems introduced at the beginning
of the chapter  where the goal is
to try to develop simple numerical estimates for various quantities of
interest by pure thought.   Not only does the Fermi approach allow
us to estimate key magnitudes, but even more importantly, it is
one of the most powerful ways I know to make sure that the
stories we tell about our data actually make sense.
   In this section, we ask ourselves whether we can understand some
of the numerical values reported throughout this chapter as well
as what key scaling results we should bear in mind when
thinking about membranes.  We pass through each of the sections
of the chapter in turn, each time taking the opportunity to reflect on
the numbers we have seen.

{\it Size and shape redux.}
In the first part of the chapter, we considered different ways of characterizing
the size and shapes of membranes and the molecules that make them up.
This led us to the fascinating experiments of Langmuir that used the relationship
between tension and area as a way of determining the size and shape of lipids.
Here, our aim is to use order-of-magnitude thinking to try
and put those numbers in perspective.
As an example from everyday life where a simple numerical estimate of
the Fermi type can  help us build intuition by giving us a sense of the relative
sizes of membranes and the cells they envelop, we consider
the fuselage of an airplane.  One of the most popular tourist destinations in
Seattle is the factory of Boeing where one can see giant airplanes such
as the 747, 777 and 787 in the process of assembly.  As part of that tour one is treated to the
view of a cross-section of a 747 fuselage which gives a sense of just how
thin the skin of an airplane really is.   For the perceptive flyer, this
same observation can be made upon entering the plane by looking
at the fuselage near the door. What one notices is that the  the exterior shell of
the plane is less than a centimeter thick while the overall diameter of that very same fuselage is roughly 5 m, resulting in an aspect ratio of 1:500.
Interestingly, the aspect ratio of cell membrane width to cell size
is quite comparable to those of an airplane fuselage. For a 2 micron cell size, typical of a bacterium,  the 4 nm thickness of its cell membrane implies a similar aspect ratio of 1:500.

%

{\it Concentrations.}
The section on concentrations reminded us that cell membranes are made up of molecules
and that even in tiny bacterial cells, there are tens of millions of such molecules
of hundreds of different types.  A very simple order of magnitude result
that emerges from these numbers is a naive estimate of the rates of
lipid synthesis.  Specifically, if the membrane area has to double
during the cell cycle, this tells us that the number of lipids in the cell membrane
has to double.  For a bacterium such as {\it E. coli},
this means that if a typical bacterium has $2 \times 10^7$ lipids and the cell cycle is roughly 2000 seconds,
then the rate of lipid synthesis is roughly
\begin{equation}
\mbox{bacterial lipid synthesis rate} = {\mbox{number of lipids} \over \mbox{cell cycle time}}
\approx {2 \times 10^7~ \mbox{lipids}  \over 2 \times 10^3~ \mbox{s}} \approx 10^4 ~\mbox{lipids/s}.
\end{equation}
It is deeply interesting to think of how the many different types of lipids are each
synthesized with the correct rates to maintain the overall concentration distribution.

 Another critical concern in our discussion of the chemistry of membranes was how to think
 about the relative abundance of lipids and proteins.   One of the interesting ways
 to broach this question is through reference to the fraction of genomes that is
 devoted to membrane proteins.  We can examine this question both from
 a genomic point of view and from a proteomic point of view.  Scientists have
 become increasingly adept at reading genomes and as a result, by recognizing
 features such as transmembrane alpha helices, it is possible to estimate
 the fraction of proteins that are membrane proteins with a rule of
 thumb being that roughly 1/4 of the protein coding genes correspond to membrane
 proteins.\cite{Fagerberg2010}
 From a proteomic point of view, this question can be addressed
 by asking what are the copy numbers of these different membrane
 proteins.    Given that a bacterium such as {\it E. coli} has several million
 proteins in total, what fraction of those proteins are in the membrane?
 To give a feeling for the answer to that question, we ask about the
 copy numbers of some key membrane proteins.   Specifically, we consider
 membrane transporters, components of the ATP synthesis machinery
 and the receptors of chemotaxis to give an idea of the molecular census for some of the most important classes of membrane proteins.  Transport of sugars across the cell membrane
 is one of the most critical activities of growing bacteria.  Recent ribosome profiling measurements and mass spectrometry measurements tell us that the number of copies
 of sugar transporters for glucose (ptsI proteins, a component of the phosphoenolpyruvate-dependent sugar phosphotransferase system) have a copy number of between roughly 3000 copies per
 cell and 15,000 copies per cell depending upon the growth conditions.\cite{Li2014,Schmidt2016}  We examine the relevance of
 these numbers in the context of membrane dynamics below.   ATP synthase is one of the most important of
 membrane protein components in almost all cell types.   In {\it E. coli}
 the ATP synthase complex is built up of many different subunits.  For those
 subunits that come with a stoichiometry
 of one molecule per complex,
 their copy number ranges  between
 3000 and 10,000 copies per cell.\cite{Li2014,Schmidt2016} Knowing these numbers provides a powerful
 sanity check on the rate of ATP production per cell since
 with roughly 3000 such synthases, each rotating at about 300 turns per second (BNID 104890), this means that over a cell cycle of 2000~s, on the order of $10^9$ ATPs will
 be generated, comparable to the number needed to run the cellular economy. \cite{Phillips2012,Milo2016}
  Finally, for the chemotaxis receptors
 such as Tar and Tsr, the copy numbers can be as low as several thousand
 and as high as nearly 40,000 per cell (BNID 100182).\cite{LiHazelbauer2004,Bitbol2015}
 These numbers give us a sense that if roughly 1000 of the 4000 or so {\it E. coli} proteins
 are membrane proteins and each comes with a copy number of roughly 1000, then
 a first simple estimate is that there are a total of $10^6$ membrane proteins distributed across the inner and
 outer membranes of these cells.


{\it Membrane mechanics.}  Our section on membrane mechanics
gave us a basis for thinking about many key processes that take place in cell
biology.  One such example that begins to shed light on the free energy demands
associated with sculpting membranes into different shapes is that of
membrane vesicles.  From the standpoint of the energetic description given
in eqn.~\ref{eqn:Helfrich2}, we can
make a simple estimate of the free energy
cost required to create spherical vesicles such as those found at
synapses.    Since for a sphere  the two radii of curvature are equal  and have a value $R$
and the total area of each such sphere is $4\pi R^2$, eqn.~\ref{eqn:Helfrich} instructs us to sum up
\begin{equation}
E_{vesicle}={\kappa_B \over 2} ({1 \over R}+{1 \over R})^2  A_{sphere}
= {\kappa_B \over 2} {4 \over R^2} 4 \pi R^2.
\end{equation}
This implies the fascinating and for many people, counterintuitive result, that the  energetic cost for vesicle formation due to membrane bending is $E_{vesicle}= 8 \pi \kappa_B \approx 250-500 ~k_BT$, completely independent of the size of the vesicle.

A second example from membrane mechanics is to try to estimate the strain
suffered by a membrane at the time of rupture.  To estimate this magnitude,
we can use
\begin{equation}
\tau_{rupture}=K_A {\Delta A_{rupture} \over A},
\end{equation}
where the subscript {\it rupture} indicates the value of the parameter
at rupture.
If we use the values provided in table~\ref{tab:1}, we can estimate the
rupture strain as
\begin{equation}
 {\Delta A_{rupture} \over A}= {\tau_{rupture} \over K_A} \approx {5~\mbox{mN/m} \over
 200~\mbox{mN/m}} \approx 2.5 \%.
 \end{equation}
 Often people are surprised by how small the rupture strains really are
 since we have an impression that lipid bilayers are floppy, squishy and highly
 deformable materials.




{\it Membrane dynamics.}
In the section on the Dynamic Membrane, we considered the
permeability of membranes to various molecular species.
One simple estimate that we can do to get a sense of the meaning
of the permeability is to ask how many molecules cross
the cell membrane each second given some concentration difference.
Given the concept provided in eqn.~\ref{eqn:Permeability}, we can
estimate
\begin{equation}
{dN \over dt}=j\times A,
\end{equation}
Given a typical membrane permeability for water of order
$p\approx 100~\mu$m/s and considering a typical concentration
difference of salt across the cell membrane when cells are subjected to  an osmotic shock  of order 100~mM $\approx
10^8~\mbox{molecules}/\mu$m$^3$, \cite{Bialecka2015}  for example, we can make the simple estimate that
\begin{equation}
j \times A = p \Delta c A \approx 100~\mu m/s \times 10^8/\mu m^3 \times 5 \mu m^2 \approx 5 \times 10^{10} s^{-1}.
\end{equation}
These numbers are interesting to contrast with the rate of transport of
molecules across ion channels.  Specifically, given the conductivity of a channel
such as the mechanosensitive channel of large conductance (MscL), we find that the opening of a single channel yields a flow rate
of several molecules per nanosecond, quite comparable to the flow rate
rate of water across the membrane itself.\cite{Louhivuori2010,Haswell2011}

One of the most interesting estimates concerning membrane dynamics that
we can consider focuses on the mass and energy economy of a cell.  To this
day, I still marvel at the fact that one can take 5~mL of liquid containing
some salts and sugars, inoculate that solution with a single bacterium, and 12 hours later
one will find as many as $10^9$ cells per mL of solution.  Effectively, what has happened
is that the molecules in the medium have been taken up by that bacterium, used
to construct building materials and energetic molecules such as ATP and then used them to construct a new cell.  This process repeats over and over again every 20 or so minutes.
These observations raise an obvious Fermi question: is the rate of membrane transport
of sugar molecules, for example, fast enough to keep up with the needs of
the cell to reproduce.\cite{Milo2016}  To approach that question, we consider the flux of
sugar across the membrane using the numbers presented above, namely,
that there are 3000-15,000 sugar transporters per cell, each of which is
able to take up sugars at a rate of several hundred sugars/sec (BNID 102931, 103160, 100736).
We can get a feeling for the number of sugars taken up per cell cycle as
\begin{equation}
\mbox{flux of sugar}=(10^4~\mbox{transporters}) \times (300~\mbox{sugars/transporter sec} )\times (2 \times 10^3~\mbox{seconds}) \approx 6 \times 10^9~\mbox{sugars}.
\end{equation}
This number is of the right order, though probably on the low side of
what is needed to power the cellular economy and raises interesting
questions about possible rate-limiting steps in cellular growth.\cite{Phillips2012,Milo2016}

 Just as we did in the section of the chapter on the Dynamic Membrane, it is of interest
 to focus not only on the dynamics across the membrane, but also on the dynamics
 of molecules within the membrane.  Specifically, one question of interest is how long does it take molecules to travel across the cell membrane given the measured diffusion
 constants?
     To answer this question, we appeal to
 the simple estimate that the time scale for diffusing a distance $L$ is given by
 \begin{equation}
 t_{diffusion} \approx {L^2 \over D}.
 \end{equation}
 For a bacterial cell with dimensions of several microns, this means that the diffusion
 time to explore the membrane is
 \begin{equation}
 t_{diffusion} \approx { 1~\mu m^2 \over 1~\mu m^2/s} \approx 1~s,
 \end{equation}
 where we have taken a diffusion coefficient for a lipid of 1~$\mu m^2/s$.
 This characteristic time scale is confirmed in fluorescence-recovery-after-photobleaching (FRAP) experiments (see \cite{Kumar2010}, for example).

{\it The Electrical Membrane.}
The electric fields across biological membranes are surprisingly high
as can be estimated by using
\begin{equation}
E \approx {V \over d} \approx {100~\mbox{mV} \over 4~\mbox{nm}} \approx
{100 \times 10^{-3}~\mbox{V} \over 4 \times 10^{-9}~\mbox{m}}
\approx 2.5 \times 10^7~{\mbox{V} \over \mbox{m}}.
\end{equation}
Note that this field is an order of magnitude higher than the electric
fields associated with dielectric breakdown in the atmosphere.  And yet,
 fields five times as high have been measured in membranes with no evidence
for any anomalous behavior.\cite{Andersen1983}


This section had as its ambition to give a sense of how the numbers
summarized in Table~\ref{tab:1} can be used to develop intuition.
\cite{Mahajan2014,Phillips2012}  In fact, more than anything,
this brief section is an invitation to others to look for meaning
in the hard won outcome of the recent work to extend
membrane numeracy.

%

\section*{The Missing Membrane Numbers}

As a final send off of this brief ode to biological numeracy
for membranes, we reflect on the state of
our art and how it can be improved.
Despite a long list of truly amazing successes, there are still many things not to like
about the current status of biological numeracy, not only in terms of
how well we actually know the numbers, but also in terms of what those numbers
might mean for a deeper understanding of biological systems.
The goal in this final section is to make an attempt at critiquing
both this article and the current state of the art with the aim
of suggesting future directions.  Though  the ``by the numbers'' approach has become something of a cliche, my opinion remains
that there is much to be gained by pushing hard with this approach on each of the many diverse
and wonderful facets of biology.\cite{Bintu2005aFinal,Garcia2010Final,Phillips2009Milo,
Moran2010,Flamholtz2014, Milo2016, Shamir2016}

One of the first weaknesses of biological numeracy in the membrane setting
(and beyond) is the need to establish measurements of sufficient precision
that we can confidently report on measured values.  For example, there is
already much evidence that biological membranes ``care'' about their lipid
composition.  It would be a powerful addition to our ability to ferret out
molecular mechanisms to be able to examine these membrane
compositions for all organelles as a function of time and for a variety of
different environmental conditions.  First steps in this direction have
been made in thinking about proteomes with one of my favorites
reporting on the proteome of {\it E. coli} in more than twenty distinct conditions.\cite{Schmidt2016}  Absent accurate and reproducible measurements in
the membrane setting,
we are handcuffed in our efforts to construct a fruitful dialogue between
theory and experiment.\cite{Phillips2015-2, Bialek2015}

A second important challenge for the future of membrane numeracy
is the vast differences between model membranes and the real world
of plasma and organellar membranes.  Effectively each and every section
of this chapter - size and shape, composition, mechanics, transport,
electrical properties - is bereft of any deep understanding of how
all of the heterogeneities of real membranes might alter
the numbers, and what the significances of such alterations might be.
The advent of mass spectrometry in conjunction
with ever more sophisticated microscopies  as a window onto membrane composition have left in their wake a host of mysteries and challenges.
As highlighted in Figures~\ref{fig:YeastLipids}  and ~\ref{fig:OrganelleLipids},
and indicated widely in other literature,\cite{Takamori2006,Dupuy2008,VanMeer2008, Ejsing2009,Kalvodova2009,Sampaio2011,Layre2011,Carvalho2012,Klose2012, Klose2013} cells care about their lipid composition.
What is lacking is a conceptual framework that tells us what
these numbers really mean in terms of biological function,
what they imply about the regulation of lipid biochemistry and perhaps most
importantly, what they imply about the evolution of life.

Another example that strikes me as an exciting challenge to our current thinking
broadly concerns the question of cellular shape, and the  shapes of organelles,
more specifically.   The images shown in Figure~\ref{fig:EukaryoticOrganelles}
make clear the great diversity of membrane shapes.  The study of
mitochondria as a concrete example presents challenges at every
turn.\cite{Neupert2012}   My personal favorite remains the intriguing
membrane structures found in the outer segments of photoreceptors
(see Figure~\ref{fig:EukaryoticOrganelles}B).  In the context of
the ideas presented in this chapter, one of the ways that people
have attacked questions of shape traditionally has been through
the approach of free energy minimization.\cite{Seifert1997,Boal}
But there are interesting, novel alternatives that are now in play.
One approach  focuses  on the role of dynamics
where there is an interplay between differential growth
and the cost of elastic deformations as characterized by the kinds of mechanical parameters reported here.\cite{Savin2011,Shyer2013}

Thus far our discussion has largely focused on the physical properties of membranes.
But there is another interesting angle on membranes that is more related to
their evolutionary significance.   Interestingly, one of
the simplest acts of biological numeracy, namely, counting, can provide evolutionary
insights.    Specifically, the number of membranes surrounding an organelle is perhaps the best indicator of its evolutionary origins, with the argument being made that more than one such membrane means that organelle has an endosymbiotic origin and more than two such membranes might imply nested
symbioses.\cite{DacksPeden2009,Field2009,Richardson2015}


We are in the midst of a biological revolution.  The pace of discovery in the study of
living matter is dizzying in all corners of biology.  The central thesis of
this article is enlightenment through biological numeracy.  That is,
as part of our attempt to make sense of the living world, we can sharpen
our questions and be more rigorous in our demands about what
it means to really understand something.\cite{Phillips2015-2, Bialek2015}  One of the ways of placing those
demands is to ask for an interplay between our experimental data
and our theoretical understanding of biological processes.  The study
of biological membranes is one of the most important areas for future work
and in many ways has not kept pace with insights into genomes and the
proteins they code for because of a want of appropriate tools.  It is hoped
that the chapters in this book will serve as an inspiration for the development
of the tools that will make membrane numeracy as sophisticated as is our
understanding of nucleic acids and proteins. \\

\noindent {\bf Acknowledgments}\\


One of the best parts of being a member of
the scientific enterprise is all the smart and
interesting people we get to interact with.  In preparing
this chapter I sent out a survey to many experts in membrane
biology and biophysics and was overwhelmed with
the thoughtful responses that I received from
 many  colleagues.
I am grateful to Olaf Andersen, Patricia Bassereau, Joel Dacks, Markus Deserno,
Evan Evans, Ben Freund, Jay Groves, Christoph Haselwandter,  Liz Haswell, KC Huang, Ron Kaback, Heun Jin Lee,  Mike Lynch, Bill Klug, Jane Kondev, Ron Milo, Uri Moran, John Nagle,
 Phil Nelson, Bert Poolman, Tom Powers,  Doug Rees,
  James Saenz, Pierre Sens,  Victor Sourjik,
Stephanie Tristram-Nagle,
    and Tristan Ursell for useful discussions.
I am especially grateful to Olaf Andersen, Markus Deserno, Christoph Haselwandter, James Saenz, Pierre Sens and
Tristan Ursell who have been patient and persistent in advancing my
membrane education, though obviously all shortcomings in this chapter
are due to my failure to absorb that education and are no fault of their own.
I am privileged to be entrusted by
the National Science Foundation, the National Institutes of Health, The California
Institute of Technology and La Fondation Pierre Gilles de Gennes with the funds that make the kind of  work described here possible.
Specifically I am grateful to the NIH for support through award numbers DP1
OD000217 (Director�s Pioneer Award),  R01 GM085286 and R01
GM084211.
I am also grateful to the Kavli Institute for Theoretical Physics where much of this chapter
was written.  More generally, this article is part of an adventure that I have undertaken
with Ron Milo and Nigel Orme (our illustrator) and generously funded by the Donna and Benjamin Rosen Bioengineering Center at Caltech.  Finally and sadly, since the completion of
this chapter, my friend and collaborator Bill Klug was brutally murdered in his office
by a former graduate student.  I had asked Bill to join me in the writing of this chapter,
but he was too busy during this summer and instead of having the happy presence of his name as a co-author I instead have the solemn and unhappy duty to dedicate this
short piece to him, kind and intellectually deep, above all a family man, he will be deeply missed.
\\



%
%
\bibliography{MWCPaper2012,PaperLibrary,PaperLibrary2015}

\end{document}